\documentclass[a4paper,11pt]{article}

\makeatletter

\usepackage{graphicx}

\usepackage{colortbl}
\usepackage{amsmath}
\usepackage{amssymb}
\usepackage{mathrsfs}
\usepackage{amsthm}
\usepackage{bm}
\usepackage{natbib}
\usepackage{titling}

\usepackage{multirow}

\usepackage[plain,noend]{algorithm2e}
\usepackage{xr}
\newtheorem{thm}{Theorem}
\newtheorem{lem}{Lemma}
\newtheorem{defi}{Definition}

\newtheorem{prop}{Proposition}

\newcommand{\sk}[1]{{\color{black}#1}}
\newcommand{\skb}[1]{{\color{black}#1}}
\newcommand{\skc}[1]{{\color{black}#1}}
\newcommand{\skd}[1]{{\color{black}#1}}
\newcommand{\ske}[1]{{\color{black}#1}}
\newcommand{\skf}[1]{{\color{black}#1}}
\newcommand{\skg}[1]{{\color{black}#1}}
\newcommand{\skh}[1]{{\color{black}#1}}

\begin{document}

\title{\textbf{\Large Copula-based measures of asymmetry between the lower and upper tail probabilities} \vspace{0.2cm} \\
}
\author{{\scshape Shogo Kato\thanks{\footnotesize {\it Address for correspondence}: Shogo Kato, Institute of Statistical Mathematics, 10-3 Midori-cho, Tachikawa, Tokyo 190-8562, Japan. {\tt E-mail:\ skato@ism.ac.jp}} \(^{,a}\), \ Toshinao Yoshiba\(^{b,c}\)  and Shinto Eguchi\,\(^{a}\)} \vspace{0.5cm}\\
\textit{\(^{a}\) Institute of Statistical Mathematics} \\  
\textit{\(^{b}\) Tokyo Metropolitan University} \\
\textit{\(^{c}\) Bank of Japan}
}
\date{August 4, 2020}
\maketitle



\begin{abstract}
We propose a copula-based measure of asymmetry between the lower and upper tail probabilities of bivariate distributions.
The proposed measure has a simple form and possesses some desirable properties as a measure of asymmetry.
The \skc{limit} of the proposed measure as the index goes to the boundary of its domain can be expressed in \skc{a simple form} under certain conditions on copulas.
\skc{A sample analogue of the proposed measure for a sample from a copula is presented and its weak convergence to a Gaussian process is shown.
\skh{Another} sample analogue of the presented measure, which is based on a sample from a distribution
 on $\mathbb{R}^2$, is given.}
Simple methods for interval estimation and nonparametric testing based on the \skc{two} sample analogues are presented.
As an example, the presented measure is applied to \skc{daily returns of S\&P500 and Nikkei225}.
\vspace{0.4cm}

\end{abstract}

\sk{
\section{Introduction}
In statistical analysis of multivariate data, it is often the case that data have complex dependence structure \skh{among variables}.
As a statistical tool for analyzing such data, copulas have gained \skh{their} popularity in various academic fields, especially, finance, actuarial science and survival analysis
\citep[see, e.g.,][]{joe97,joe14,nel,mcn}.

A copula is a multivariate cumulative distribution function with uniform $[0,1]$ margins.
The bivariate case of Sklar's theorem states that, for a bivariate cumulative distribution function $F$ with margins $F_1$ and $F_2$, there exists a copula $C$ such that $F(x_1,x_2)=C(F_1(x_1),F_2(x_2))$.
Hence a copula can be used as a model for dependence structure and is applicable for flexible modeling.
Another important advantage of using copulas is that copulas are useful as \skh{measures} of dependence.
For example, the tail dependence coefficient is well-known as a measure of dependence in a tail \citep[see, e.g., Section 2.13 of][]{joe14}. 

One important problem in copula-based modeling is to decide which should be fitted to data of interest, a copula with symmetric tails or a copula with asymmetric tails. 
An additional problem arising from this is that if a copula with asymmetric tails is appropriate for the data, how much degree of tail asymmetry the copula should have.
These problems are important because the lack of fit in tails of copulas leads to erroneous results in statistical analysis.
For example, it is said that widespread applications of Gaussian copula, which has symmetric light tails, to financial products have contributed to the global financial crisis of 2008--2009 \citep[see][]{don}. 
\skd{Therefore}, in order to carry out decent statistical analysis, it is essential to evaluate the degree of tail asymmetry of copula appropriately.
Given the stock market spooked by the outbreak of COVID-19, these problems would be even more important.

Some copula-based measures of tail asymmetry have been proposed in the literature.
\citet{nik} and \citet{dob} discussed a measure of tail asymmetry based on the difference between the conditional Spearman's rhos for truncated data.
\skh{\citet{kru17} proposed an extension of their measure, which can regulate weights of tails.
\citet{ros} proposed three measures of tail asymmetry;
two of them are based on moments or quantiles of a transformed univariate random variable and one of them is based on a difference between a copula and its reflected copula.
As related works, measures of radial symmetry for the entire domain, not for tails, have been proposed, for example, by \citet{deh} and \citet{gen}.
See \citet[Section 2.14]{joe14} for the book treatment on this topic.}

In this paper we propose a new copula-based measure of asymmetry between the lower and upper tails of bivariate distributions.
The proposed measure and its sample analogues have various tractable properties;
the proposed measure has a simple form and its calculation is fast;
the proposed measure possesses \skh{desirable properties which as} a measure of tail asymmetry;
the limits of the proposed measure as the index goes to the boundary of its domain can be easily evaluated under mild conditions on copulas;
sample analogues of the proposed measure converge weakly to a Gaussian process or its mixture;
simple methods for interval estimation and hypothesis testing based on the sample analogues are available;
a multivariate extension of the proposed measure is straightforward.

The paper is organized as follows.
In Section \ref{sec:definition} we propose a new copula-based measure of \skh{tail} asymmetry and present its basic properties.
Section \ref{sec:limits} considers the \skh{limits} of our measure as the index goes to the boundary of its domain.
Values of the proposed measure for some well-known copulas are discussed in Section \ref{sec:existing}.
In Section \ref{sec:estimator} two sample analogues of the proposed measure are presented and their asymptotic properties are investigated.
Also statistical inference for our measure such as interval estimation and hypothesis tests is discussed and a simulation study is carried out to demonstrate the results.
In Section \ref{sec:comparison} the proposed measure is compared with \skh{other} copula-based measures of tail asymmetry.
In Section \ref{sec:example} the proposed measure is applied to daily returns of S\&P500 and Nikkei225.
Finally, a multivariate extension of the proposed measure is briefly considered in Section \ref{sec:discussion}.

Throughout this paper, a `copula' refers to the bivariate case of a copula, namely, a bivariate cumulative distribution function with uniform $[0,1]$ margins.
Let $\mathscr{C}$ be a set of all the bivariate copulas.
Let $\overline{C}$ denote the survival copula associated with $C$, which defined by $\overline{C}(u_1,u_2) = 1 - u_1 - u_2 + C(u_1,u_2)$.
Define $\bar{u}$ by $\bar{u}=1-u$.
}

\section{Definition and basic properties} \label{sec:definition}
In this section we propose a measure for comparing the probabilities of the lower and upper tails of bivariate distributions.
The proposed measure is defined as follows.
\begin{defi} \label{def:alpha}
Let $(X_1,X_2)$ be an $\mathbb{R}^2$-valued random vector.
Assume $X_1$ and $X_2$ have continuous margins $F_1$ and $F_2$, respectively.
Then a measure of comparison between the lower-left and \skh{upper-right} tail probabilities of $(X_1,X_2)$ is defined by
$$
\alpha (u) = \log \left( \frac{ \mathbb{P}(F_1(X_1) > 1-u , F_2(X_2) > 1- u)}{\mathbb{P}(F_1(X_1) \leq u , F_2(X_2) \leq u)} \right), \quad 0 < u \leq 0.5. \label{eq:alpha}
$$
Here the definition of the logarithm function is extended to be $\log (x/y) = -\infty$ if $x=0$ and $y>0$, $\log (x/y) = \infty $ if $x>0$ and $y=0$, and $\log (x/y)=0$ if $x=y=0$.
\end{defi}

Similarly it is possible to define a measure \skh{to compare} the lower-right and upper-left tail probabilities of bivariate distributions.
Properties of this measure immediately follow from those of $\alpha (u)$, \sk{which will be given hereafter}, by replacing $(X_1,X_2)$ by $(X_1,-X_2)$.

The calculation of $\alpha (u)$ can be simplified if the distribution of $(X_1,X_2)$ is represented in terms of copula.
The proof is straightforward and therefore omitted.
\skd{\begin{prop} \label{thm:copula}
Let $C$ denote \skc{a} copula of $(X_1,X_2)$ given by $C(u_1,u_2) = \mathbb{P}(F_1$\\
$(X_1) \leq u_1, F_2(X_2) \leq u_2)$.
Then $\alpha (u)$ \skc{defined in Definition \ref{def:alpha}} can be expressed as
\begin{equation}
\alpha (u) = \log \left( \frac{2u -1 + C(1-u,1-u)}{C(u,u)} \right). \label{eq:alpha}
\end{equation}
\end{prop}}
Note that, using the survival copula $\overline{C}$ associated with $C$ \skc{with $\bar{u} =1-u$}, the proposed measure (\ref{eq:alpha}) has the simpler expression
$$
\alpha (u) =  \log \left( \frac{\overline{C}(\bar{u},\bar{u})}{C(u,u)} \right).
$$


\sk{
Throughout this paper, the lower $[0,u]^2$ tail and the upper $[1-u,1]^2$ tail of the copula $C$ are said to be \textit{symmetric} if $C(u,u) = \overline{C}(\bar{u},\bar{u})$.

Unlike many existing measures, the proposed measure (\ref{eq:alpha}) is not a global measure but a local one in the sense that this measure focuses on the probability of a \skh{subdomain of the copula regulated} by the index $u$.
Setting a particular value of $u$ or looking at the behavior of $\alpha(u)$ for multiple choices of $u$, the proposed measure (\ref{eq:alpha}) provides a different insight from the global measure.
For more details on the comparison between the proposed measure and existing ones, see Section \ref{sec:comparison}.
}

It is straightforward to see that the following basic properties hold for $\alpha (u)$.
\skd{ \begin{prop} \label{prop:basic}
Let $\mathscr{C}$ be a set of all bivariate copulas.
Denote the measure $\alpha (u)$ for the copula $C \in \mathscr{C}$ by $\alpha_C (u)$.
Assume that $p_L=C(u,u)$, $p_U=\overline{C} (\bar{u},\bar{u})$, and $C_P(u,v) = C(v,u)$ is the permutated copula of $C$.
Then, for $0<u \leq 0.5$, we have that:
\begin{itemize}
\item[(i)] $- \infty \leq \alpha_C (u) \leq \infty$ for every $C \in \mathscr{C}$; the equality holds only when either $p_U=0$ or $p_L=0$;
\item[(ii)] $\alpha_C (u) = 0$ if and only if $p_L=p_U$;
\item[(iii)] for fixed $p_U$, $\alpha_C(u)$ is monotonically non-increasing with respect to $p_L$;
similarly, for fixed $p_L$, $\alpha_C(u)$ is monotonically non-decreasing with respect to $p_U$;
\item[(iv)] $\alpha_C (u) = - \alpha_{\overline{C}} (u)$ for every $C \in \mathscr{C}$;
\item[(v)] $\alpha_{C_P}(u) = \alpha_{C}(u)$ for every $C \in \mathscr{C}$;
\item[(vi)] if $C \in \mathscr{C}$ and $\{C_n\}_{n \in \mathbb{N}}$ is a sequence of copulas such that $C_n \rightarrow C$ uniformly, then $\alpha(C_n) \rightarrow \alpha (C)$.
\end{itemize}
\end{prop} }

\sk{
Property (i) implies that the proposed measure is potentially unbounded \skh{although it is bounded except for the unusual case $p_U=0$ or $p_L=0$.
Compared with a similar measure based on the difference between $p_U$ and $p_L$, our measure is advantageous in the sensitivity of detecting the asymmetry of tail probabilities for small $u$; see Section \ref{sec:alternative} for details}.
\skh{Property (ii) implies} that $\alpha_C(u) =0$ for any $0 < u \leq 0.5$ if the copula $C$ is \skg{radially} symmetric, namely, $C \equiv \overline{C}$.
Property (ii) is the same as an axiom of \citet{deh} and is an extended property of \citet{ros}.
Properties (iv)--(vi) are the same as the axioms of tail asymmetry presented in Section 2 of \skh{\citet{ros}.
\skh{It is possible to use any function of $p_U/p_L$ other than the logarithm function as a measure of tail asymmetry.
However one nice property of the proposed measure is property (iv) which other functions of $p_U/p_L$ do not have in general.}
}
}


\section{Limits of the proposed measure} \label{sec:limits}

\sk{We consider limits of the proposed measure (\ref{eq:alpha}) \skh{the index goes to the boundary of its domain}.
It follows from the expression (\ref{eq:alpha}) that
$\alpha (0.5) = 0,$
for any copula $C \in \mathscr{C}$.
Therefore we have
\begin{equation}
\lim_{u \uparrow 0.5} \alpha (u) = 0.  \label{eq:alpha05}
\end{equation}
The limiting behavior of $\alpha (u)$ as $u \rightarrow 0$ is much more intricate.
To consider this problem, define
\begin{equation}
\alpha(0) = \lim_{u \downarrow 0} \alpha (u), \label{eq:alpha0}
\end{equation}
given that the limit exists.}
\skh{Here we present three expressions for the limit (\ref{eq:alpha0}).

\skh{The first expression} is based on the tail dependence coefficients. }
Tail dependence coefficients are often used as local dependence measures of bivariate distributions.
The lower-left and upper-right tail dependence coefficients of the random variables $X_1$ and $X_2$ are defined by
$$
\lambda_L = \lim_{u \downarrow 0} \frac{ \mathbb{P} (F_1(X_1) \leq u, F_2(X_2) \leq u)}{u} \label{eq:lambda_l} \quad
\mbox{and} \quad
\lambda_U = \lim_{u \uparrow 1} \frac{ \mathbb{P} (F_1(X_1) > u, F_2(X_2) > u)}{1-u}, \label{eq:lambda_u}
$$
respectively, given the limits exist.
If $(X_1,X_2)$ has the copula $C$, the expressions for $\lambda_L$ and $\lambda_U$ are simplified as
\begin{equation}
\lambda_L = \lim_{u \downarrow 0} \frac{C(u,u)}{u} \quad \mbox{and} \quad \lambda_U = \lim_{u \uparrow 1} \frac{\overline{C}(u,u)}{1-u}, \label{eq:lambda_u2}
\end{equation}
respectively \citep[see, e.g.,][Section 2.13]{joe14}.\begin{thm} \label{thm:tail_dep}
Let $(X_1,X_2)$ be an $\mathbb{R}^2$-valued random vector with the copula $C$.
Assume that the lower-left and upper-right tail dependence coefficients of $X_1$ and $X_2$ exist and are given by $\lambda_L$ and $\lambda_U$, respectively.
Suppose that either $\lambda_L$ or $\lambda_U$ is not equal to zero.
Then
$$
\alpha(0) = \log \left( \frac{\lambda_U}{\lambda_L} \right).
$$
\end{thm}
\skd{See Supplementary Material for the proof.}
\sk{Theorem \ref{thm:tail_dep} can be generalized by utilizing the concepts of tail orders and tail order parameters.
If there exists a constant $\kappa_L > 0$ and a slowly varying function $\ell_L (u)$ such that
$ C(u,u) \sim u^{\skc{\kappa_L}} \ell_L (u)$  \skh{$(u \rightarrow 0)$},
then $\kappa_L$ is called the lower tail order of $C$ and $\Upsilon_L=\lim_{u \downarrow 0} \ell_L (u)$ is called the lower tail order parameter of $C$\skc{, where $f(u) \sim g(u)$ $ (u \rightarrow 0)$ is defined by $\lim_{u \downarrow 0} f(u)/g(u)=1$.}
Similarly, the upper tail order and the upper tail order parameter of $C$ are defined by the lower tail order and the lower tail order parameter of the survival copula $\overline{C}$, respectively.
See \citet[Section 2.16]{joe14} for more details on the tail orders and tail order parameters.
Using the tail orders and tail order parameters, we have the following result.
\skb{The proof is given in Supplementary Material.}
\begin{thm}  \label{thm:tail_order}
Let $\kappa_L$ and $\kappa_U$ be the lower and upper tail orders of the copula $C$, respectively.
Then $\alpha (0) = \infty$ if $\kappa_L < \kappa_U$ and $\alpha (0) = -\infty$ if $\kappa_L > \kappa_U$.
If $\kappa_L=\kappa_U$ and either of the lower tail order parameter $\Upsilon_L$ or the upper tail order parameter $\Upsilon_U$ of $C$ is not equal to zero, then
$\alpha (0) = \skc{\log (\Upsilon_U / \Upsilon_L )}$.
\end{thm}
Note that Theorem \ref{thm:tail_order} with $\kappa_L=\kappa_U=1$ reduces to Theorem \ref{thm:tail_dep}.
Theorems \ref{thm:tail_dep} and \ref{thm:tail_order} are useful to evaluate $\alpha(0)$ if we already know \skh{the} tail dependence coefficients or tail orders and tail order parameters of a copula.
If those values are not known, the following \skh{third expression for $\alpha(0)$} could be useful.
}
\begin{thm} \label{thm:alpha0_den}
\sk{
Let $(X_1,X_2)$ be an $\mathbb{R}^2$-valued random vector with the copula $C$.
Suppose that there exists $\varepsilon >0$ such that $c(u) = d^2 C(t,t)/ dt^2 |_{t=u}$ exists in $(0,\varepsilon) \cup (1-\varepsilon,1)$.
Assume that $\lim_{u \downarrow 0} d C(u,u) /du = \lim_{u \downarrow 0} d \overline{C}(\bar{u},\bar{u}) / du  = 0$.
Then
$$
\alpha(0) = \log \left(  \lim_{u \downarrow 0} \frac{c(1-u)}{c(u)} \right),
$$
given the limit exists.}
\end{thm}
\skd{See Supplementary Material for the proof.}
\sk{
As will be seen in the next section, Theorem \ref{thm:alpha0_den} can be utilized to calculate $\alpha(0)$ for Clayton copula and Ali-Mikhail-Haq copula.
}

\section{Values of the proposed measure for some existing copulas} \label{sec:existing}

\skh{In this section we discuss the values of the proposed measure $\alpha (u)$ for some existing copulas.
It is seen to be useful to plot $\alpha (u)$ with respect to $u$ for comparing the probabilities of the lower $[0,u]^2$ tail and upper $[1-u,1]^2$ one for the whole range of $u \in (0,0.5]$.}
See, e.g., \citet{joe14} for the definitions of the existing copulas discussed in this section.
\vspace{0.2cm}

\subsection{Copulas with symmetric tails}
\skd{Proposition \ref{prop:basic}} implies that $\alpha(u)=0$ for any $u \in [0,0.5]$ if $C(u,u)=\overline{C}(\bar{u},\bar{u})$ for any $u \in [0,0.5]$.
Such copulas include the independence copula, Gaussian copula, $t$-copula, Plackett copula and FGM copula.
\skh{Among well-known Archimedean copulas, Frank copula has a radially symmetric shape} and therefore $\alpha(u)=0$ for any $u$.

\subsection{Copulas with asymmetric tails} \label{sec:asymmetric_copula}
There exist various copulas for which $\alpha (u)$ is not equal to zero in general.
Many Archimedean copulas have asymmetric \skh{tails}, including Clayton copula, Gumbel copula, Ali-Mikhail-Haq copula and two-parameter BB copulas.
In addition, some asymmetric extensions of Gaussian copula and $t$-copula have been proposed recently.
Such \skh{extensions} include \skh{the skew-normal copulas and skew-$t$ copulas discussed in \citet{joe06} and \citet{yos},} for which $\alpha(u)$ is not equal to zero in general.
As examples of copulas with asymmetric \skh{tails}, here we discuss the values of $\alpha(u)$ for \skh{the three well-known copulas, namely, Clayton copula, Ali-Mikhail-Haq copula and BB7 copula.} \vspace{0.2cm}

\noindent \textit{Clayton copula:}
Clayton copula is defined by
\begin{equation}
C_{cl} (u_1,u_2; \theta) = \max \left\{ u_1^{-\theta} + u_2^{-\theta} - 1 ,0 \right\}^{-1/\theta}, \label{eq:clayton}
\end{equation}
where $\theta \in [-1,\infty) \setminus \{0\}$.
\begin{figure}
\begin{center}
\begin{tabular}{ccc}
\includegraphics[width=4.5cm,height=4.5cm]{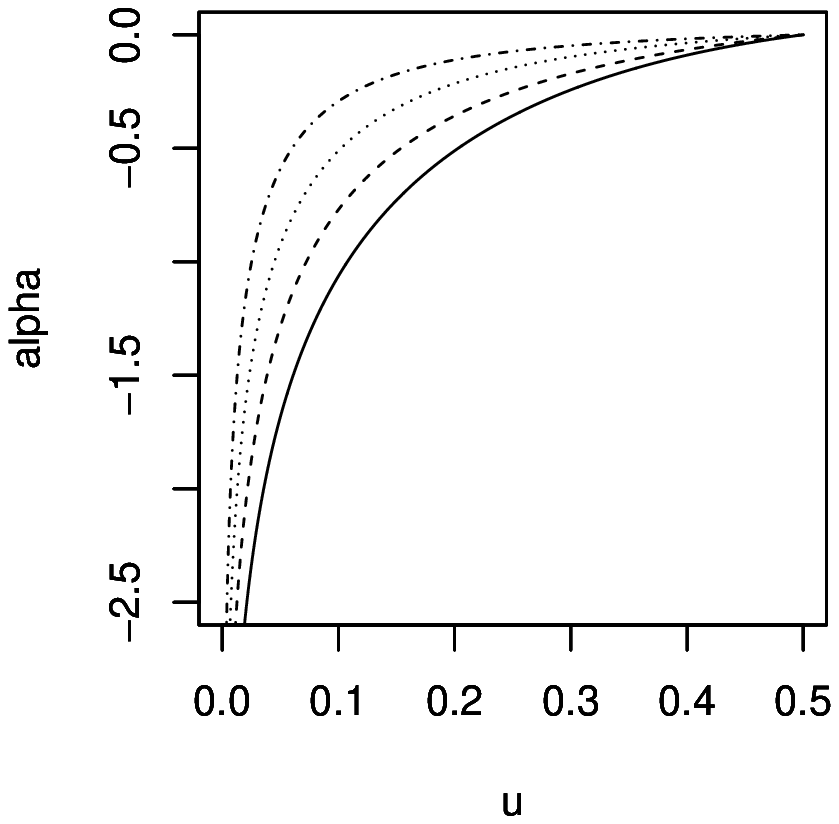} & 
\includegraphics[width=4.5cm,height=4.5cm]{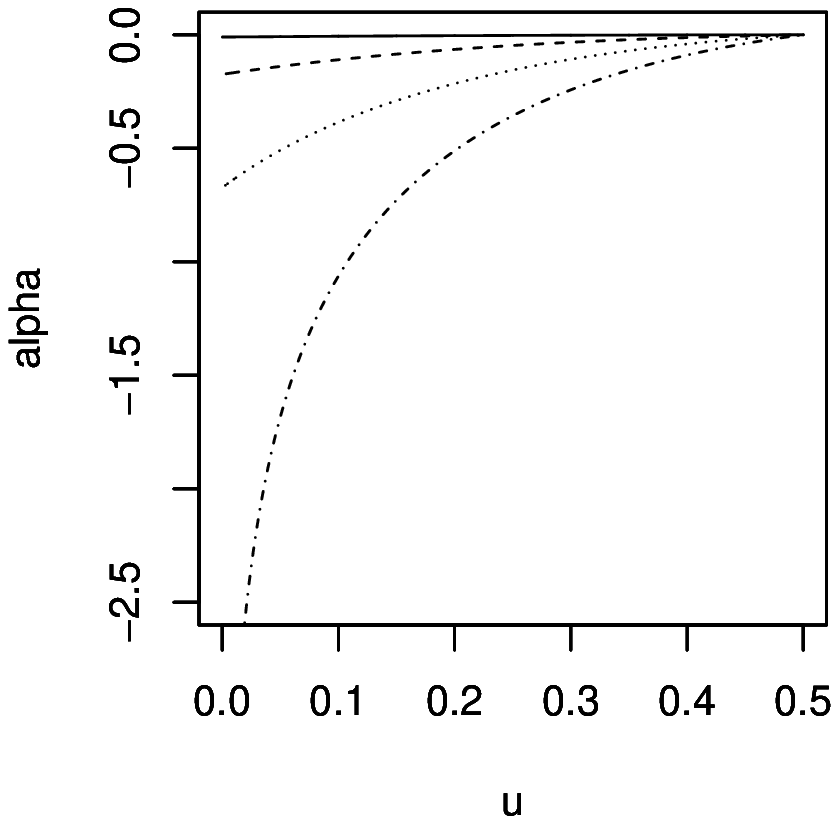} &
\includegraphics[width=4.5cm,height=4.5cm]{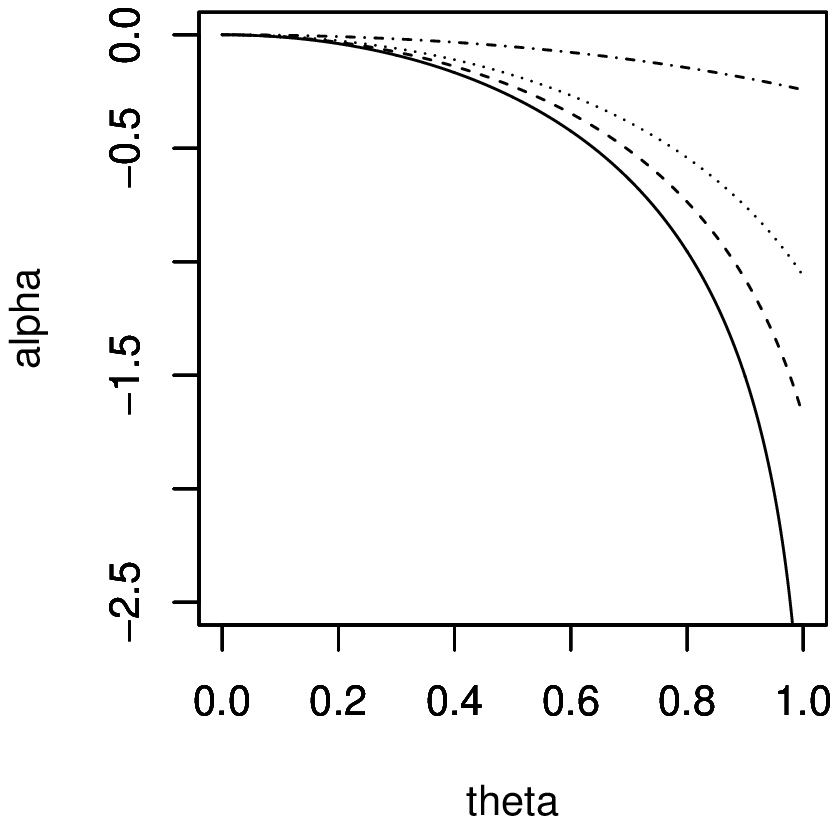} \\ 
\hspace{0.9cm} (a) & \hspace{0.9cm} (b) & \hspace{0.9cm} (c)
\end{tabular}
\caption{Plots of $\alpha(u)$ for: (a) Clayton copula (\ref{eq:clayton}) with respect to $u$ for $\theta=1$ (solid), $\theta=5$ (dashed), $\theta=10$ (dotted), and $\theta=20$ (dotdashed), (b) Ali-Mikhail-Haq copula (\ref{eq:amh}) with respect to $u$ for $\theta=0.1$ (solid), $\theta=0.4$ (dashed), $\theta=0.7$ (dotted), and $\theta=1$ (dotdashed), and (c) Ali-Mikhail-Haq copula (\ref{eq:amh}) with respect to $\theta$ for $u=0.01$ (solid), $u=0.05$ (dashed), $u=0.1$ (dotted), and $u=0.3$ (dotdashed).
} \label{fig:alpha_clay_amh}
\end{center}
\end{figure}
Figure \ref{fig:alpha_clay_amh}(a) plots the values of $\alpha(u)$ as a function of $u$ for four positive values of $\theta$.
\skb{(For an intuitive understanding of the distributions of Clayton copula, see Figure \ref{fig:rv}(a) and (b) of \skd{Supplementary Material} which plot random variates from Clayton copula with the two values of the parameters used in Figure \ref{fig:alpha_clay_amh}.)}
As is clear from equation (\ref{eq:alpha05}), $\alpha(0.5)=0$ for any $\theta$.
The smaller the value of $u$, the smaller the value of $\alpha(u)$.
The figure also suggests that, for a fixed value of $u$, as $\theta$ increases, the value of $\alpha (u)$ approaches zero.
The upper tail dependence coefficient of Clayton copula is 0 \skh{and the} lower tail dependence coefficient is $2^{-1/\theta}$ for $\theta >0$ and $0$ for $\theta \leq 0$.
Therefore, for $\theta > 0$, Theorem \ref{thm:tail_dep} implies that $\alpha(0) = -\infty$, meaning that the lower tail dependence is considerably stronger than the upper \skh{one}.
If $\theta \in [-1,0)$, it follows from Theorem \ref{thm:alpha0_den} that $\alpha(0) = \infty$. \vspace{0.2cm}

\noindent \textit{Ali-Mikhail-Haq copula:}
Ali-Mikhail-Haq copula is of the form
\begin{equation}
C(u_1,u_2) = \frac{u_1 u_2}{1-\theta (1-u_1)(1-u_2)},  \label{eq:amh}
\end{equation}
where $\theta \in [-1,1]$.
The values of $\alpha(u)$ as a function of $u$ and $\theta$ are exhibited in Figure \ref{fig:alpha_clay_amh}(b) and (c), respectively.
(See Figure \ref{fig:rv}(d) and (e) of \skd{Supplementary Material} for plots of random variates generated from Ali-Mikhail-Haq copula with the two values of the parameters used in Figure \ref{fig:alpha_clay_amh}(b).)
Figure \ref{fig:alpha_clay_amh}(b) suggests that \skh{$\alpha(u)$ decreases with $u$}.
Also it appears that, for a fixed value of $u$, the greater the value of $\theta$, the smaller the value of $\alpha(u)$.
This observation can be seen more clearly in Figure \ref{fig:alpha_clay_amh}(c) which plots the values of $\alpha(u)$ as a function of $\theta$.
Since both the lower and upper tail dependence coefficients of this copula are equal to zero, one can not apply Theorem \ref{thm:tail_dep} for the calculation of $\alpha(0)$.
However Theorems \ref{thm:tail_order} and \ref{thm:alpha0_den} are applicable in this case and we have a simple form $\alpha(0) = \log (1-\theta^2)$.
\vspace{0.2cm}

\noindent \textit{BB7 copula:}
Finally, consider the BB7 copula of \citet{joe96} defined by
\begin{equation}
C(u_1,u_2) = 1 - \left[ 1- \left\{ \left( 1-\overline{u}_1^\theta \right)^{-\delta} + \left( 1-\overline{u}_2^\theta \right)^{-\delta} -1 \right\}^{-1/\delta} \right]^{1/\theta}, \label{eq:bb7}
\end{equation}
where $\delta >0$ and $\theta \geq 1$.
Unlike the last two copulas, this model has two parameters.
The parameter $\delta$ controls the lower tail dependence coefficient, while $\theta$ regulates the upper \skh{one}.
Indeed, the lower and upper tail dependence coefficients are known to be $2^{-1/\delta}$ and $2-2^{1/\theta}$, respectively.

It follows from Theorem \ref{thm:tail_dep} that $\alpha(0)=\log(2-2^{1/\theta} ) - \delta^{-1} \log 2$.
\begin{figure}
\begin{center}
\begin{tabular}{cc}
\includegraphics[width=6cm,height=6cm]{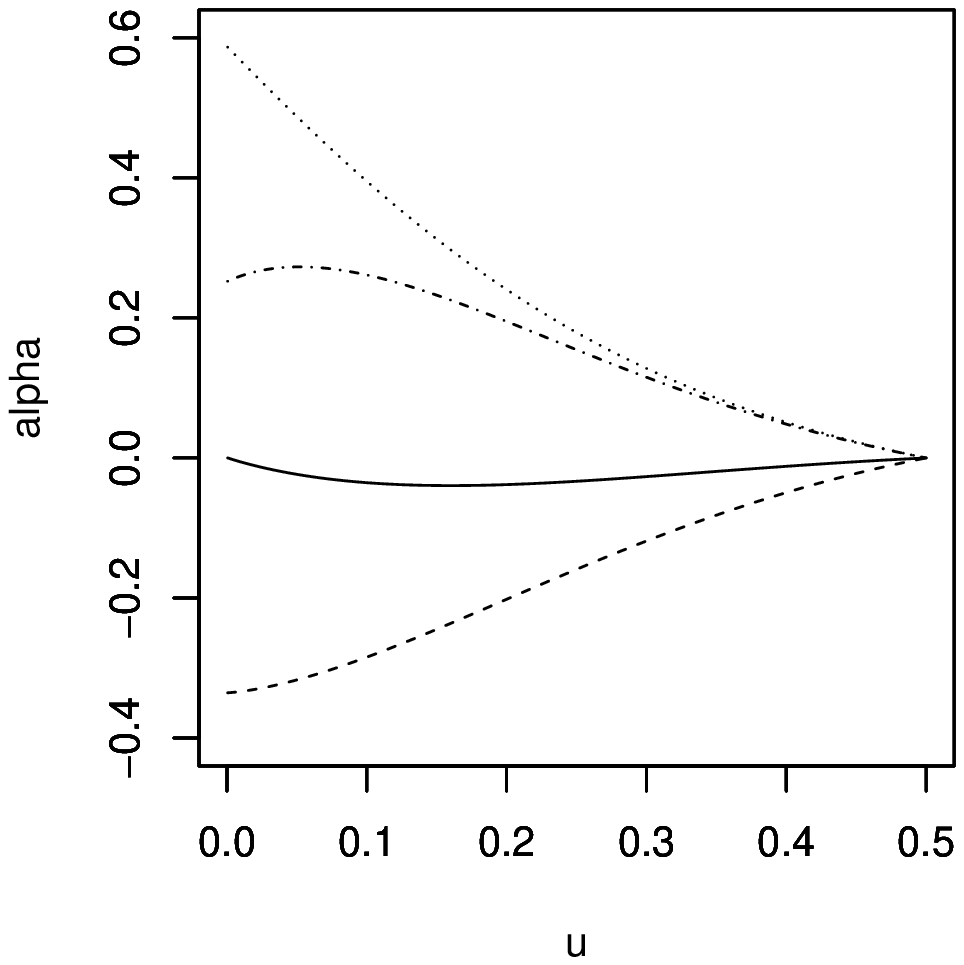} \quad & \quad
\includegraphics[width=6cm,height=6cm]{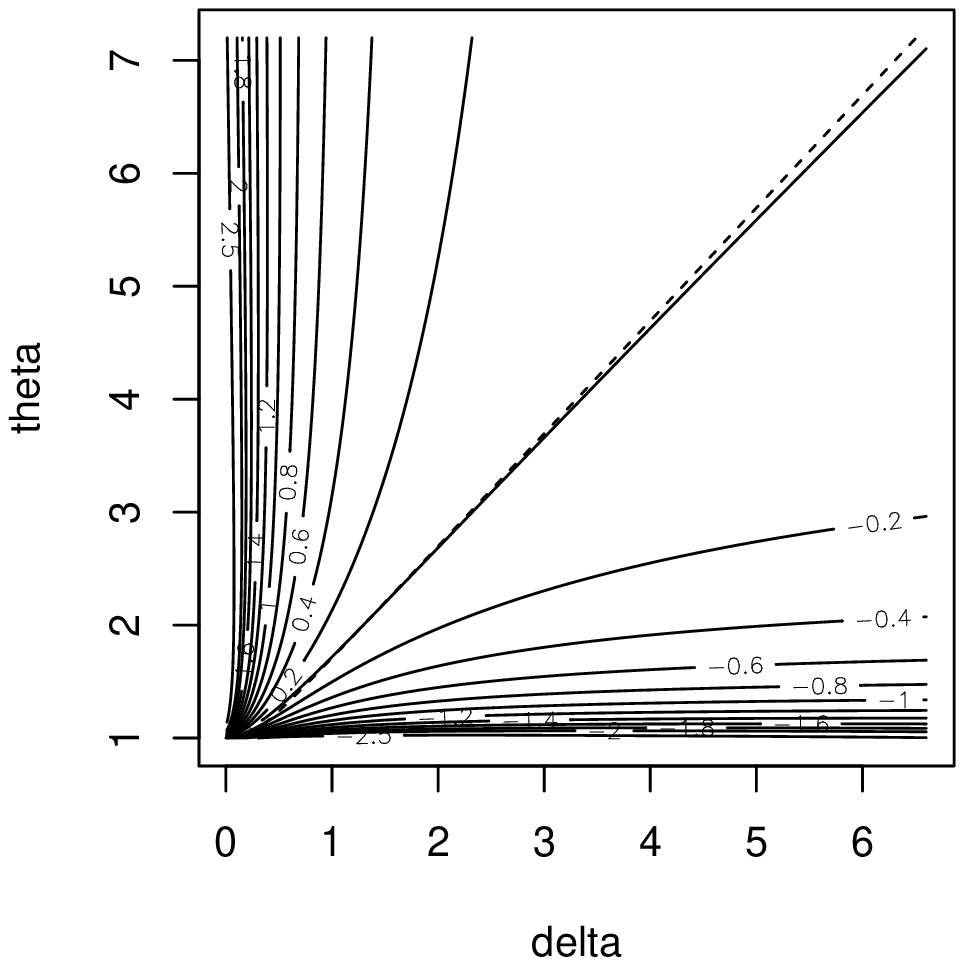} \\ 
\hspace{0.9cm} (a) & \hspace{1.5cm} (b)
\end{tabular}
\caption{(a) Plot of $\alpha(u)$ for BB7 copula (\ref{eq:bb7}) with respect to $u$ for $(\delta,\theta)=(1,1.71)$ (solid), $(\delta,\theta)=(1.94,1.71)$ (dashed), $(\delta,\theta)=(1,7.27)$ (dotted), and $(\delta,\theta)=(1.94,7.27)$ (dotdashed).
(b) Contour plot of $\alpha(0.01)$ (solid) and plot of $\alpha(0)=0$ (dashed) with respect to $(\delta,\theta)$.
} \label{fig:alpha_bb7}
\end{center}
\end{figure}
Figure \ref{fig:alpha_bb7} displays a plot of $\alpha(u)$ with respect to $u$ for four selected values of $(\delta,\theta)$ and that of $\alpha(u)$ with respect to $(\delta,\theta)$ for $u=0.01$.
\skh{Note that, in} Figure \ref{fig:alpha_bb7}(a), $\delta=1$ and $\delta=1.94$ imply that the lower tail dependence coefficients are around 0.5 and 0.9, respectively, while $\theta=1.71$ and $\theta=7.27$ suggest that the upper tail dependence coefficients are about 0.5 and 0.7, respectively.
\skb{(See also Figure \ref{fig:rv}(g)--(h) of \skd{Supplementary Material} for plots of random variates from BB7 copula (\ref{eq:bb7}) with the three combinations of the parameters in Figure \ref{fig:alpha_bb7}(a).)}
Figure \ref{fig:alpha_bb7}(a) suggests that, when both the lower and upper tail dependence coefficients are around 0.5, the values of $\alpha(u)$ are close to zero for any $u$.
When the difference between the lower and upper tail dependence coefficients is large, $\alpha(u)$ appears to be monotonic with respect to $u$.
It can be \skh{seen} from Figure \ref{fig:alpha_bb7}(b) that the values of $\alpha(0.01)$ monotonically decreases as $\delta$ increases.
Also, \skh{$\alpha(0.01)$ monotonically increases with $\theta$}.
The two \skh{contours} $\alpha(0.01)=0$ and $\alpha(0)=0$ show somewhat similar \skh{shapes}, implying that $\alpha(0.01)=0$ is a reasonable approximation to $\alpha(0)=0$.

\section{Two sample analogues of $\alpha(u)$} \label{sec:estimator}

\sk{
In practice, it is often the case that the form of the copula $C(u_1,u_2)$ underlying data is not known.
In such a case, we need to estimate $\alpha(u)$ based on the data.
In Sections \ref{sec:alpha_sample_def}--\ref{sec:test} we propose a sample analogue of $\alpha (u)$ based on a sample from the copula.
Section \ref{sec:alpha_empirical} presents a sample analogue of $\alpha(u)$ based on a sample from a distribution on $\mathbb{R}^2$.
A comparison between the two proposed sample analogues of $\alpha(u)$ is discussed via a simulation study in Section \ref{sec:simulation}.
}

\subsection{A sample analogue of $\alpha(u)$ based on a sample from a copula} \label{sec:alpha_sample_def}

A sample analogue of $\alpha(u)$ based on a sample from a copula is defined as follows.
\begin{defi} \label{defi:hat_alpha}
Let $(U_{11},U_{21}), \ldots, (U_{1n},U_{2n})$ be a random sample from a copula.
Then we define a sample analogue of $\alpha(u)$ by
$$
\hat{\alpha}(u) = \log \left( \frac{T_U(u)}{T_L(u)} \right), \label{eq:alpha_hat}
$$
where
$$
T_L(u) = \frac{1}{n} \sum_{i=1}^n \bm{1} (U_{1i} \leq u, U_{2i} \leq u), \label{eq:t_l}
$$
$$
T_U(u) = \frac{1}{n} \sum_{i=1}^n \bm{1} (U_{1i} \geq 1-u, U_{2i} \geq 1-u ), \label{eq:t_u}
$$
and $\bm{1}(\cdot)$ is an indicator function, i.e., $\bm{1}(A)=1$ if $A$ is true and $\bm{1}(A)=0$ otherwise.
\end{defi}
In Sections \ref{sec:alpha_sample_def}--\ref{sec:test}, we assume that $(U_{11},U_{21}),$$\ldots,$$(U_{1n},U_{2n})$ is an iid sample from the copula $C(u_1,u_2)$.
\sk{For iid $\mathbb{R}^2$-valued random vectors $(X_{11},X_{21})$$,\ldots,$ $(X_{1n},X_{2n})$, if the margins of $X_{11}$ and $X_{21}$ are known to be $F_1$ and $F_2$, respectively, then $\hat{\alpha}(u)$ can be obtained by replacing $(U_{1j},U_{2j})$ by $(F_1(X_{1j}),F_2(X_{2j}))$ $(j=1,\ldots,n)$.}

The goal of this subsection is to investigate some properties of $\hat{\alpha}(u)$.
To achieve this, we first show the following lemma.
\skb{
See \skd{Supplementary Material} for the proof.
\begin{lem} \label{lem:t}
For $0 < u,v \leq 0.5,$ we have the following:
$$
\mathbb{E} \left[ T_L(u) \right] = C_u, \quad
\mathbb{E} \left[ T_U(u) \right] = \overline{C}_{\bar{u}}, \quad
{\rm var} \left[ T_L (u) \right] = \frac1n C_u ( 1 - C_u ),
$$
$$
{\rm var} \left[ T_U (u) \right] = \frac1n \overline{C}_{\bar{u}} ( 1 - \overline{C}_{\bar{u}} ), \quad {\rm cov} \left[ T_L(u), T_L(v) \right] = \frac1n C_{u\wedge v} ( 1 - C_{u \vee v} ),
$$
$$
{\rm cov} \left[ T_L(u), T_U(v) \right] = -\frac1n C_{u\wedge v} \overline{C}_{\bar{u} \wedge \bar{v}}, \quad
{\rm cov} \left[ T_U(u), T_U(v) \right] = \frac1n \overline{C}_{\bar{u} \vee \bar{v}} ( 1- \overline{C}_{\bar{u} \wedge \bar{v}} ),
$$
where $u \wedge v = \min (u,v)$, $u \vee v = \max (u,v)$, and $C_w = C(w,w)$.
\end{lem}
}
This lemma implies that $\hat{\alpha}(u)$ is a consistent estimator of $\alpha(u)$.
Applying this lemma, we obtain the following asymptotic result.
\skb{The proof is given in \skd{Supplementary Material}.}

\begin{thm} \label{thm:gauss_con}
Define
$$
\mathbb{A}_n(u) = \sqrt{n} \left\{ \hat{\alpha}(u) - \alpha (u) \right\}, \quad 0 < u \leq 0.5.
$$
Then, as $n \rightarrow \infty$, $\{ \mathbb{A}_n(u) \, | \, 0 < u \leq 0.5 \}$ converges weakly to a centered Gaussian process with covariance function
\begin{equation} \skb{
\sigma (u,v) \equiv \mathbb{E} [ \mathbb{A}_n(u) \mathbb{A}_n(v) ] = \frac{C(u \vee v, u \vee v )+ \overline{C}(\bar{u} \wedge \bar{v}, \bar{u} \wedge \bar{v})}{C(u \vee v,u \vee v) \cdot \overline{C}(\bar{u} \wedge \bar{v}, \bar{u} \wedge \bar{v})}. \label{eq:cov}
} \end{equation}
\end{thm}

\skc{We note that} the covariance function (\ref{eq:cov}) is monotonically decreasing with respect to $u \vee v$.
If $u \vee v =0.5$, then the covariance function (\ref{eq:cov}) reaches the minimum value $2/C(0.5,0.5)$.



\subsection{Interval estimation based on $\hat{\alpha}(u)$} \label{sec:confidence}

\skh{An asymptotic interval estimator of $\alpha(u)$ can be obtained by applying the asymptotic results obtained in the previous subsection.
Theorem \ref{thm:gauss_con} implies that, for fixed $u \in (0, 0.5]$,
$$
\sqrt{n} \left\{ \hat{\alpha}(u) - \alpha (u) \right\} \xrightarrow{d} N(0,\sigma^2(u) ) \quad (n \rightarrow \infty),
$$
where $\sigma^2 (u)=\sigma (u,u)$ and $\sigma(u,u)$ is defined as in equation (\ref{eq:cov}).
Since $\sigma(u)$ includes the \skh{copula} $C$ which is usually not known in practice, we use an estimator of $\sigma (u)$ defined by
$$
\hat{\sigma} (u) = \sqrt{ \frac{T_L(u) + T_U(u)}{T_L(u) \cdot T_U(u)}}.
$$
It follows from Lemma \ref{lem:t} that
$
\hat{\sigma} (u) \xrightarrow{a.s.} \sigma (u)
$
as $n \rightarrow \infty$.
Then we have
$
\sqrt{n} \{ \hat{\alpha}(u) - \alpha (u) \} / \hat{\sigma} (u) \xrightarrow{d} N(0,1)
$
as $n \rightarrow \infty$.
Hence a $100(1-p)$\% nonparametric asymptotic confidence interval for $\alpha(u)$ is
\begin{equation}
\hat{\alpha}(u) - \frac{ z_{p/2} \hat{\sigma} (u) }{\sqrt{n}} \leq \alpha (u) \leq \hat{\alpha}(u) + \frac{ z_{p/2} \hat{\sigma} (u)  }{\sqrt{n}}, \label{eq:interval}
\end{equation}
where $z_{p/2}$ satisfies $\mathbb{P}(Z \geq z_{p/2}) =p/2$, where $Z \sim N(0,1)$ and $0 < p < 1$.
}

Notice that the asymptotic confidence interval (\ref{eq:interval}) is a pointwise one \sk{for a fixed value of $u$}.
If the interest of statistical analysis is to construct an asymptotic confidence band for $\alpha(u)$ for a range of $u$, one can adopt Bonferroni correction.
This can be done by replacing $p$ by $p/n$ in the equation (\ref{eq:interval}) for $u \in \mathbb{U}$, where $\mathbb{U}= \{ \min ( \max (u_{1i},u_{2i}) , \max (1-u_{1i},1-u_{2i}) ) \}_{i=1}^n$.
Therefore the asymptotic confidence band for $\alpha(u)$ with Bonferroni correction is
\begin{equation}
\left\{ \left[ \hat{\alpha}(u) - \frac{ z_{p/(2n)} \hat{\sigma} (u) }{\sqrt{n}} , \hat{\alpha}(u) + \frac{ z_{p/(2n)} \hat{\sigma}(u)}{\sqrt{n}} \right] \, \Biggl| \, u_{\min} \leq u \leq u_{\max} \right\}, \label{eq:bonferroni}
\end{equation}
where $u_{\min} = \min \mathbb{U} $ and $u_{\max} = \max \mathbb{U}$.
Since $\hat{\alpha}(u)$ and $\hat{\sigma} (u) $ are step functions, it suffices to evaluate the bounds of confidence intervals only for $u \in \mathbb{U}$.

\subsection{Hypothesis testing based on $\hat{\alpha}(u)$} \label{sec:test}
\sk{
Some hypothesis tests can be established based on $\hat{\alpha}(u)$.
For a given value of $u=u_0$, one can carry out a hypothesis test to test $H_0:\alpha(u_0)=\alpha_0$ against $H_1: \alpha(u_0) \neq \alpha_0$ by using the asymptotic confidence interval (\ref{eq:interval}).
Similarly, an one-sided test for the alternative hypothesis $H_1: \alpha (u_0) >\alpha_0$ or $H_1: \alpha(u_0) < \alpha_0$ can be derived by modifying the asymptotic confidence interval (\ref{eq:interval}).

\skc{If} the interest of analysis is to evaluate the values of $\hat{\alpha} (u)$ for multiple values of $u$, one can consider the test $H_0: \alpha(u) = \alpha_0 (u) $ against $H_1: \alpha(u) \neq \alpha_0(u)$ for \skc{$\{ u ; u = u_1,\ldots,u_m\}$}.
One example of such tests is based on a asymptotic confidence band based on Bonferroni correction (\ref{eq:bonferroni}).
However, since the confidence band based on Bonferroni correction is known to be conservative, especially, for dependent hypotheses, the test based on Bonferroni correction (\ref{eq:bonferroni}) is not powerful in general.
Alternatively, the following result can be used to present a test for the multiple values of $u$.
\skb{See \skd{Supplementary Material} for the proof.}
}
\begin{thm} \label{thm:test}
Let $\bm{a}= \sqrt{n} \{ \hat{\alpha}(u_1) - \alpha(u_1), \ldots, \hat{\alpha}(u_m) - \alpha(u_m) \}^T$ and $u_1 < \cdots < u_m$.
Suppose
$$
\hat{\bm \Sigma} = \left(
    \begin{array}{cccc}
      \hat{\sigma}^2(u_1) & \hat{\sigma} (u_1,u_2) & \ldots & \hat{\sigma} (u_1,u_m) \\
      \hat{\sigma} (u_1,u_2) & \hat{\sigma}^2 (u_2) & \ldots & \hat{\sigma} (u_2,u_m) \\
      \vdots & \vdots & \ddots & \vdots \\
      \hat{\sigma} (u_1,u_m) & \hat{\sigma} (u_2,u_m) & \ldots & \hat{\sigma}^2 (u_m)
    \end{array}
  \right),
$$
where
$ \hat{\sigma}^2 (u_i) = \hat{\sigma} (u_i,u_i) $ and $
\hat{\sigma} (u_i,u_j) = \{T_L(u_j) + T_U(u_j) \}/\{T_L(u_j)  T_U(u_j)\}$ $(i \leq j)$.
Assume that $\hat{\bm \Sigma}$ is \sk{invertible}.
Then
$$
\bm{a}^T \hat{\bm \Sigma}^{-1} \bm{a} \xrightarrow{d} \chi^2(m) \quad \mbox{as } n \rightarrow \infty,
$$
where $\chi^2(m)$ denotes the chi-squared distribution with $m$ degrees of freedom.
\end{thm}

Substituting $\alpha_0(u)$ into $\alpha(u)$ in the test statistic $\bm{a}^T \hat{\bm \Sigma}^{-1} \bm{a}$, the null hypothesis $\alpha(u)=\alpha_0(u)$ is rejected for a large value of $\bm{a}^T \hat{\bm \Sigma}^{-1} \bm{a}$.
In order that $\hat{\bm \Sigma}$ becomes invertible, the values of $(u_1,\ldots,u_m)$ need to be selected such that $T_U(u_i) < T_U(u_{i+1})$ and/or $T_L(u_i) < T_L(u_{i+1})$ hold for any $1 \leq i \leq m-1$. 

\subsection{A sample analogue of $\alpha (u)$ based on a sample from a distribution on $\mathbb{R}^2$} \label{sec:alpha_empirical}

The sample analogue of $\alpha (u)$ given in Definition \ref{defi:hat_alpha} can be calculated on the assumption that the \skh{margins} of the $\mathbb{R}^2$-valued random vector are known.
Here we discuss the case in which margins are unknown and empirical distributions are adopted as the margins.

\begin{defi} \label{defi:hat_alpha_star}
Let $(X_{11},X_{21}), \ldots, (X_{1n},X_{2n})$ be $\mathbb{R}^2$-valued random vectors.
Then we define a sample analogue of $\alpha(u)$ by
$$
\hat{\alpha}^* (u) = \log \left( \frac{T^*_U(u)}{T^*_L(u)} \right), \label{eq:alpha_star}
$$
where
$$
T^*_L(u) = \frac{1}{n} \sum_{i=1}^n \bm{1} ( \hat{F}_1(X_{1i}) \leq u, \hat{F}_2 (X_{2i}) \leq u ), \label{eq:t_l_emp}
$$
$$
T^*_U(u) = \frac{1}{n} \sum_{i=1}^n \bm{1} ( \hat{F}_1(X_{1i}) \geq 1-u, \hat{F}_2(X_{2i}) \geq 1-u ), \label{eq:t_u_emp}
$$
\begin{equation}
\hat{F}_j (X_{ji}) =  \frac{1}{n+1} \skc{ \sum_{k=1}^n } \bm{1}  ( X_{jk} \leq X_{ji} ), \quad j=1,2. \label{eq:f_emp} 
\end{equation}
\end{defi}
Note that the denominator of the empirical distribution function (\ref{eq:f_emp}) is defined by $n+1$ rather than $n$ in order to avoid positive bias of \skh{$\hat{\alpha}(u)$.}

The authors have not yet obtained the asymptotic distribution for $\hat{\alpha}^*(u)$.
However the following results are available regarding $T_U^*(u)$ and $T_L^*(u)$.
The proof is straightforward from \citet{fer}, \citet{tsu} and \citet{seg} and therefore omitted.
\begin{prop}
Let $(X_{11},X_{21}),\ldots,(X_{1n},X_{2n})$ be iid random vectors with the copula $C(u,v)$ and \skh{the} continuous margins.
Assume that $C(u,v)$ is differentiable with continuous $i$-th partial derivatives $(i=1,2)$.
Then, as $n \rightarrow \infty$,
$$
\sqrt{n} \left\{ T^*_L(u) - C(u,u) \right\} \xrightarrow{d} D^C (u),
$$
$$
\sqrt{n} \left\{ T^*_U(u) - \overline{C}(\bar{u},\bar{u}) \right\} \xrightarrow{d} D^{\overline{C}}(\bar{u}),
$$
where
$$
D^C(u) = U(u,u) - \frac{ \partial C(u_1,u)}{\partial u_1} \biggr|_{u_1=u} U(u,1) -  \frac{ \partial C(u,u_2)}{\partial u_2} \biggr|_{u_2=u} U(1,u),
$$
$U$ is a centered Gaussian process with covariance function
$$
\mathbb{E} [U(u_1,u_2) U(v_1,v_2) ] = C( u_1 \wedge v_1, u_2 \wedge v_2 ) - C(u_1,u_2) C(v_1,v_2),
$$
and $u \wedge v$ is defined as in Lemma \ref{lem:t}.
\end{prop}

Confidence intervals for $\alpha^*(u)$ can be numerically constructed using the bootstrap method.
\sk{Hypothesis tests can also be established based on the bootstrap confidence intervals.}
It should be noted that, in order to calculate $\alpha^*(u)$ based on bootstrap samples, $\hat{F}_1$ and $\hat{F}_2$ in (\ref{eq:f_emp}) should be calculated based on each bootstrap sample.
If $\hat{F}_1$ and $\hat{F}_2$ are calculated from the original data, the bootstrap confidence intervals become similar to the asymptotic confidence intervals (\ref{eq:interval}) for large $n$.

\subsection{Simulation study} \label{sec:simulation}

In order to compare the performance of the two proposed sample analogues of $\alpha (u)$ for a large sample size, we consider the following cumulative distribution function
\begin{equation}
F(x_1,x_2) = C_{cl}(F_1(x_1),F_2(x_2); 20), \quad -\infty < x_1,x_2 < \infty, \label{eq:clayton_cauchy}
\end{equation}
where $F_j(x)$ is the cumulative distribution function of the standard Cauchy distribution, i.e., $F_j(x) = 0.5+ \pi^{-1} \arctan x$, and $C_{cl}(u_1,u_2;\theta)$ denotes the Clayton copula (\ref{eq:clayton}).

\begin{figure}
\begin{center}
\includegraphics[width=12cm,height=7.2cm]{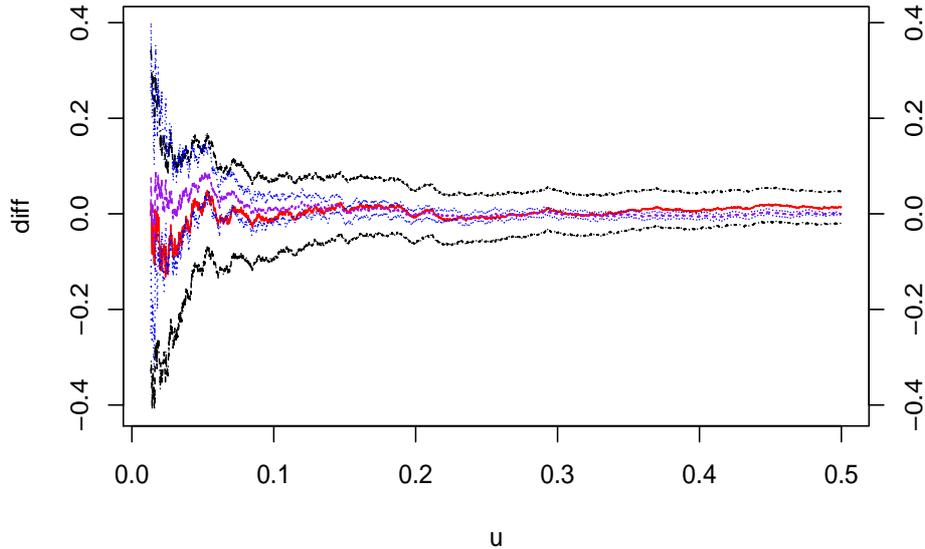} 
\end{center}
\caption{Plot of $\hat{\alpha}(u)-\alpha(u)$ \skh{(solid, red)}, $\hat{\alpha}^*(u) - \alpha(u)$ \skh{(dashed, purple)}, the lower and upper bounds of 90$\%$ asymptotic confidence intervals of $\hat{\alpha}(u) - \alpha(u)$ \skh{(dotdashed, black)}, and the lower and upper bounds of 90\% bootstrap confidence intervals $\hat{\alpha}^*(u)-\alpha(u)$ \skh{(dotted, blue)} obtained from a sample of size 10000 from the distribution (\ref{eq:clayton_cauchy}).} \label{fig:sample}
\end{figure}

Figure \ref{fig:sample} plots the values of $\hat{\alpha}(u)-\alpha(u)$, $\hat{\alpha}^*(u) - \alpha(u)$ and their \skh{bounds of} 90\% confidence intervals for a sample of size $n=10000$ from the distribution (\ref{eq:clayton_cauchy}).
See also Figure \ref{fig:alpha_clay_amh}(a) for the plot of $\alpha (u)$.
For the calculations of $\hat{\alpha}(u)$ and its confidence intervals (\ref{eq:interval}), the sample $\{(x_{1i},x_{2i})\}$ is transformed into the copula sample $\{(u_{1i},u_{2i})\}$ via $u_{ji} = F_j(x_{ji})$, where $F_j$ is the true margin $(j=1,2)$.
The confidence intervals of $\alpha^*(u)$ are calculated using the basic bootstrap method based on \skc{$999$} resamples of size $10000$; see the last paragraph of Section \ref{sec:alpha_empirical} for details.
\sk{The minimum value of $u$ in the plot is defined as $u_{\rm min} = \min \{ u \in (0,0.5] ;  T_L(u) ,T_U(u), T_L^*(u),$ $T_U^*(u) \geq 30\} \simeq 0.01$ in order that the asymptotic theory is applicable.}

The figure suggests that when $u$ is around 0.2 or greater, the performance of both $\hat{\alpha}(u)$ and $\hat{\alpha}^*(u)$ seems satisfactory.
For $u \leq 0.2$, \skh{the difference between the sample analogues and the true value increases with $u$} in general.
It appears that the 90\% confidence intervals of both $\hat{\alpha}(u)$ and $\hat{\alpha}^*(u)$ are generally narrow if $u$ is around 0.2 or more.
For $u \leq 0.2$, the smaller the value of $u$, the wider the ranges of the confidence intervals of both $\hat{\alpha}(u)$ and $\hat{\alpha}^*(u)$.
\skh{Interestingly, the} confidence intervals of $\hat{\alpha}^*(u)$ \skh{are} narrower than those of $\hat{\alpha}(u)$ in most of the plotted range of $u$.
This tendency is particularly obvious in the range $[0.2,0.5]$, where the confidence intervals of $\hat{\alpha}^*(u)$ are \skh{much narrower} than those of $\hat{\alpha}(u)$.



\section{Comparison with other measures} \label{sec:comparison}

\subsection{Comparison with existing measures}

In this section we compare our measure with \skh{other copula-based measures} of tail asymmetry.
\citet{ros} proposed three measures of tail asymmetry.
One of their measures based on the distance between a copula $C$ and its survival copula is defined by
\begin{equation}
\varsigma_3 = \sup_{(u_1,u_2) \in [0,1]^2 } \left\{ \left| C(u_1,u_2) - \overline{C}(\bar{u}_1,\bar{u}_2) \right| \right\}.  \label{eq:rosco_joe}
\end{equation}
This measure has also been proposed by \citet{deh} as a limiting case of a measure of radial asymmetry for bivariate random \skh{variables}.

Our measure (\ref{eq:alpha}) has some similarities to and differences from the measure (\ref{eq:rosco_joe}).
Similarities include that both are \skh{functions of} a copula $C$ and its survival function.
Also, both measures satisfy Properties (ii), (v) and (vi) of \skd{Proposition \ref{prop:basic}}.

However there are considerable differences between the two measures (\ref{eq:alpha}) and (\ref{eq:rosco_joe}).
First, the domains of a \skf{copula the two measures evaluate} are different.
The measure (\ref{eq:rosco_joe}) is a global measure in the sense that the whole domain of the copula is taken into account to \skf{evaluate} the value of the measure, while our measure (\ref{eq:alpha}) is a local measure which focuses on squared subdomains of the copula.
By choosing the value of the index $u$, our measure (\ref{eq:alpha}) enables us to choose the subdomain of a copula which analysts are interested in.
However the prescription for selecting the value of $u$ is not always straightforward and the choice of the index $u$ could influence the results of analysis.
The index-free measure (\ref{eq:rosco_joe}) does not have such a problem.
However the supremum value of this measure is not necessarily attained in the tails of the distribution and the value of the measure might not reflect the tail probabilities \skf{if $u_1 \geq 0.5$ or $u_2 \geq 0.5$}.
\skg{Also, because of its locality, computations associated with our measure (\ref{eq:alpha}) are very fast.}

Also there are differences between the two measures (\ref{eq:alpha}) and (\ref{eq:rosco_joe}) in terms of properties.
Our measure (\ref{eq:alpha}) satisfies all the properties of (i)--(vi) of \skd{Proposition \ref{prop:basic}} which include four (out of five) axioms of \citet{ros}.
However this measure does not satisfy one of the axioms, i.e., axiom (i), of \citet{ros} and therefore the value of the measure could be unbounded \skh{for special cases}.
The measure (\ref{eq:rosco_joe}) also satisfies four axioms of \citet{ros} including the axiom (i).
On the other hand, the measure (\ref{eq:rosco_joe}) does not satisfy their axiom (iii) which is equivalent to Property (iv) of \skd{Proposition \ref{prop:basic}}, implying that the measure (\ref{eq:rosco_joe}) does not distinguish which tail probability is greater than the other one.

The other two measures of \citet{ros} are derived through different approaches.
For a bivariate random vector $(U_1,U_2)$ from a copula, the two measures are based on the moments or quantile function of the univariate random variable $U_1+U_2-1$.
Therefore these measures are \skh{essentially} different from ours which is based on the joint distribution of the bivariate random vector $(U_1,U_2)$.

Another copula-based measure for tail asymmetry has been proposed by \citet{kru17}.
It is defined by
\begin{equation}
\varrho_{K} (a,u) = \varrho_L (a,u) -  \varrho_U (a,u), \label{eq:krupskii}
\end{equation}
where $0<u \leq 0.5$, $a$ is a weighting function,
$$
\varrho_L (a,u) = {\rm cor} \left[ \left. a \left( 1 - \frac{U_1}{u} \right), a \left( 1 - \frac{U_2}{u} \right) \right| U_1 < u , U_2 < u \right],
$$
$$
\varrho_U (a,u) = {\rm cor} \left[ \left. a \left( 1 - \frac{1-U_1}{u} \right), a \left( 1 - \frac{1-U_2}{u} \right) \right| U_1 > 1-u , U_2 > 1-u \right] \skc{.}
$$
If $a(x)=x$, the measure (\ref{eq:krupskii}) reduces to the measure discussed by \citet{nik} and \citet{dob}.
Properties of each term of the measure (\ref{eq:krupskii}) have been investigated by \citet{kru15}.

The measure (\ref{eq:krupskii}) is related to ours in the sense that the values of their measures are calculated from the subdomain of a copula indexed by \skh{the} truncation parameter.
However the measure (\ref{eq:krupskii}) is based on Spearman's rhos or correlation coefficients of a truncated copula, and therefore the interpretation of the values of the measure (\ref{eq:krupskii}) is essentially different from ours.
A nice property of the measure (\ref{eq:krupskii}) is that the weights of tails can be controlled through the weight function $a$.
Therefore this measure can be a useful measure of tail asymmetry if the weight function is appropriately defined.

\subsection{An alternative measure based on tail probabilities} \label{sec:alternative}

The proposed measure $\alpha (u)$ is a function of the lower and upper tail probabilities.
Here we briefly consider another measure of comparison between the two tail probabilities.

\begin{defi}
Let $(X_1,X_2)$, $F_1$ and $F_2$ be defined as in Definition \ref{def:alpha}.
Then we define a measure of comparison between the lower-left and upper-right tail probabilities of $(X_1,X_2)$ by
\begin{align*}
\beta (u) = & \ u^{-\kappa} \bigl\{ \mathbb{P} (F_1(X_1) > 1-u , F_2(X_2) > 1- u) \nonumber \\
& - \mathbb{P} (F_1(X_1) \leq u , F_2(X_2) \leq u) \bigr\}, \quad  0 < u \leq 0.5, \label{eq:beta}
\end{align*}
where the index $\kappa$ is given by $ \kappa \geq 1$.
\end{defi}
If $(X_1,X_2)$ has the copula $C$, the expression of $\beta(u)$ can be simplified to
$$
\beta (u) = \frac{\overline{C}( \skc{\bar{u}, \bar{u}}) - C(u,u)}{u^{\kappa}}. \label{eq:beta2}
$$
Hence this measure is based on the difference between the lower and upper tail probabilities of the copula as well as the value of $\kappa$ which could be decided based on the lower and upper tail orders.

The measure $\beta(u)$ is another simple measure to compare the lower-right and upper-left tail probabilities.
Properties \skc{(ii)--(vi)} of \skd{Proposition \ref{prop:basic}} hold for $\beta(u)$.
As for the range of the measure $\beta(u)$ related to Property (i), it can be seen $-1 \leq \beta (u) \leq 1$ if $\kappa=1$ and $-\infty \leq \beta(u) \leq \infty$ otherwise.

One big difference between $\alpha(u)$ and $\beta(u)$ is that, when considering the values of the measures for $u \simeq 0$, $\alpha(u)$ and $\beta(u)$ possibly lead to different conclusions.
As an example of this, consider the Clayton copula (\ref{eq:clayton}) with $\theta<0$, which is said to have asymmetric tails \skb{(see Figure \ref{fig:rv} of \skd{Supplementary Material} for a plot of random variates from Clayton copula with $\theta=-0.3$).}
For this model, the measure $\alpha(u)$ for $u \simeq 0$ takes large positive values because $\alpha(0) = \infty$.
However, the values of $\beta(u)$ with $\kappa=1$ for $u \simeq 0$ are close to zero since $\beta(0) = 0$.
This fact about $\beta(u)$ is reasonable in one sense, but one might argue that $\beta(u)$ does not capture the asymmetry of tail probabilities appropriately.
Although this problem can be solved by selecting a different value of $\kappa$, the selection of $\kappa$, which influences the conclusion of analysis, appears difficult in practical situations where $C$ is unknown.

\skb{
\section{Example} \label{sec:example}
As an application of the proposed \skh{measure}, we consider \skc{a dataset of daily returns of two stock indices.
The dataset is taken from historical data in Yahoo Finance, available at \skc{\texttt{https://finance.yahoo.com/quote/\%5EGSPC/history/} and \texttt{https://finance.yahoo.}\\
\texttt{com/quote/\%5EN225/history/}}.
We consider stock daily returns of S\&P500 and Nikkei225 observed from the 1st of April, 2008 until the 31st of March, 2019, inclusive.}
We fit the autoregressive-generalized autoregressive conditional heteroscedastic model AR(1)-GARCH(1,1) to each of the stock daily returns using \texttt{ugarchfit} in `rugarch' package in R \citep{rco,gha}.
The Student $t$-distribution is used as the conditional density for the innovations.
We consider the residuals $\{(x_{1i} , x_{2i}) \}_{i=1}^n$ $(n=2605)$ of the fitted AR(1)-GARCH(1,1), where $x_{1i}$ and $x_{2i}$ are the residuals of S\&P500 and Nikkei225, respectively.
\ske{The residuals show unexpected changes in daily return which are not explained by the model;
if the joint plunging probability is higher than the joint soaring probability, then the proposed measures $\alpha (u)$ is supposed to be negative.}

We discuss $\hat{\alpha}(u)$ defined in Definition \ref{def:alpha} and $\hat{\alpha}^*(u)$ defined in Definition \ref{defi:hat_alpha_star}.
In order to obtain the copula sample $\{ (u_{1i},u_{2i}) \} $ for $\hat{\alpha}(u)$, we transform the residuals $\{(x_{1i},x_{2i})\}$ via $(u_{1i},u_{2i}) = (F_1 (x_{1i}) , F_2(x_{2i}))$, where $F_1$ and $F_2$ are the cumulative distribution functions of Student $t$-distribution estimated using the maximum likelihood method.
We assume, though not mathematically precise, that $F_1$ and $F_2$ are known.
\begin{figure}
\begin{center}
\begin{tabular}{ccc}
\includegraphics[width=4.3cm,height=4.3cm] {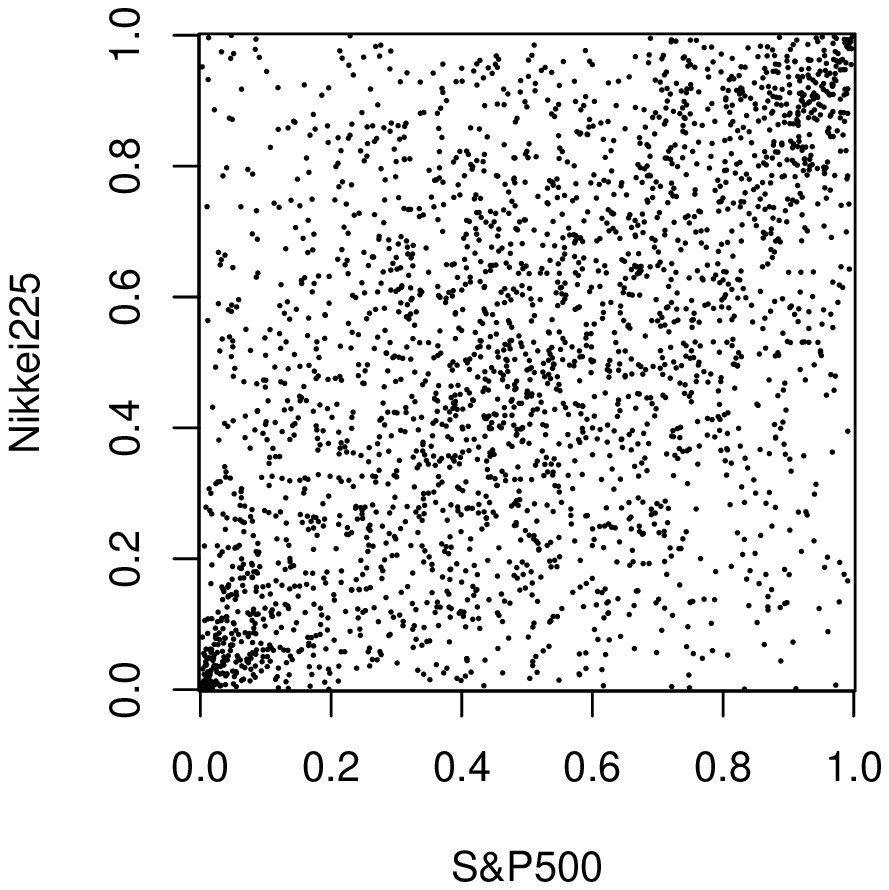} & 
\includegraphics[width=4.3cm,height=4.3cm]{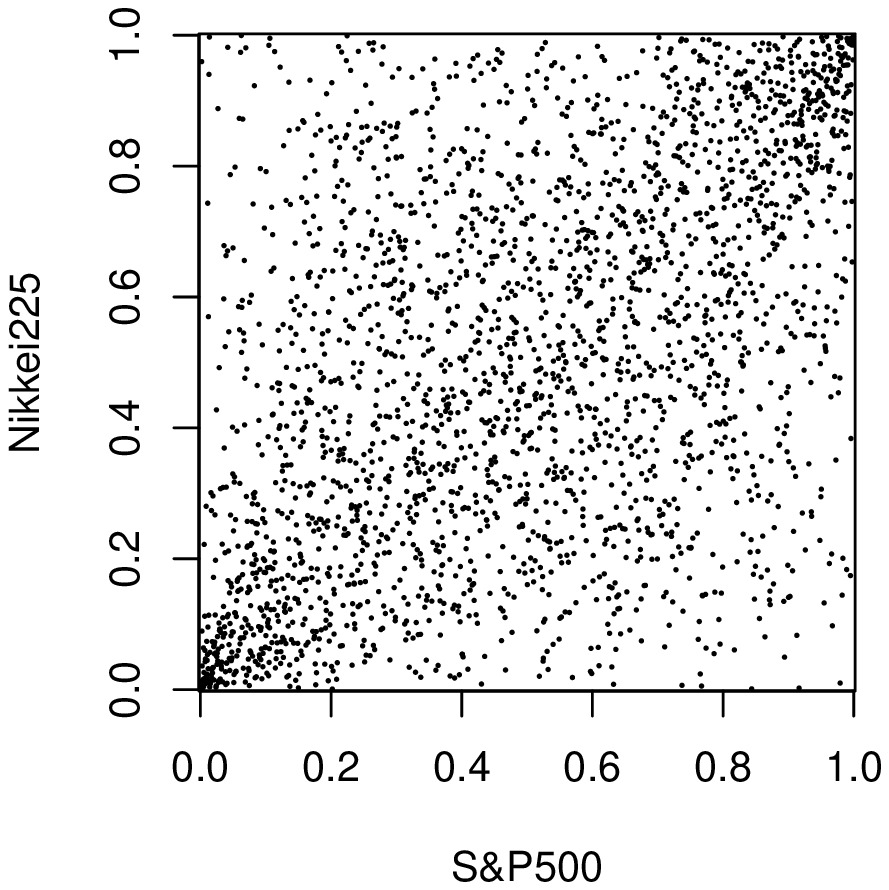} &
\includegraphics[width=4.8cm,height=4.2cm]{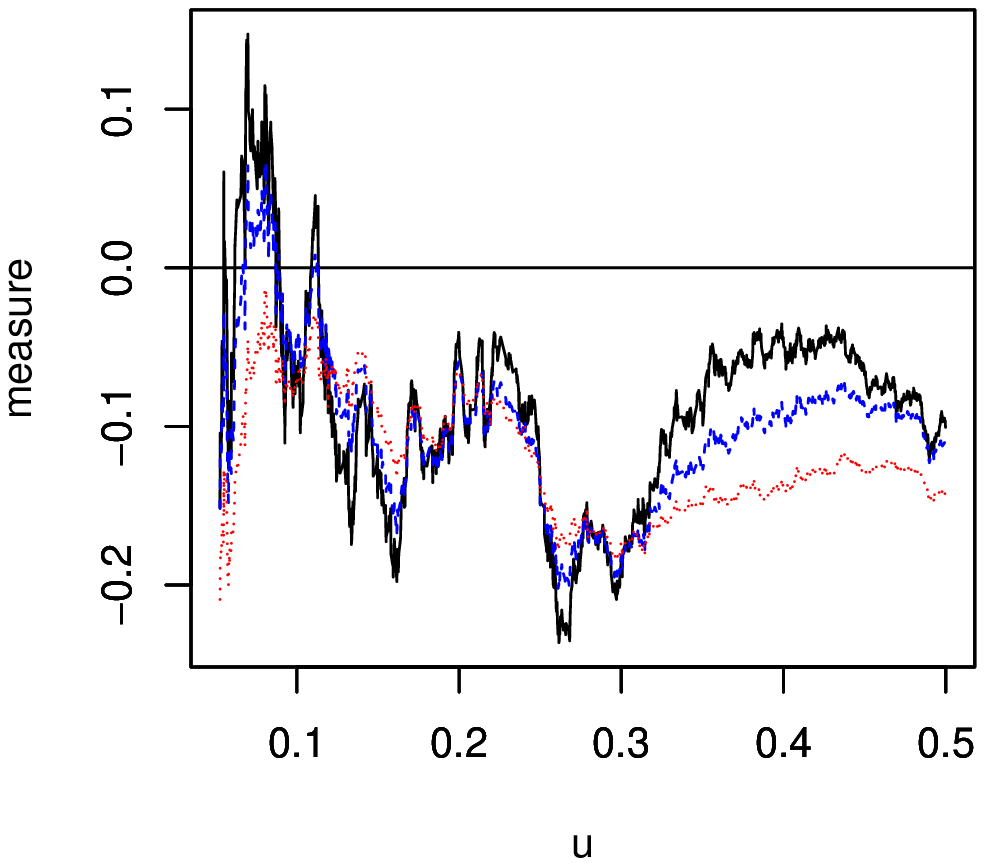} \\ 
\hspace{0.7cm} (a) & \hspace{0.7cm} (b) & \hspace{0.95cm} (c) \vspace{1.3cm}\\
\multicolumn{3}{c}{
\begin{tabular}{cc}
\includegraphics[width=7cm,height=4.97cm]{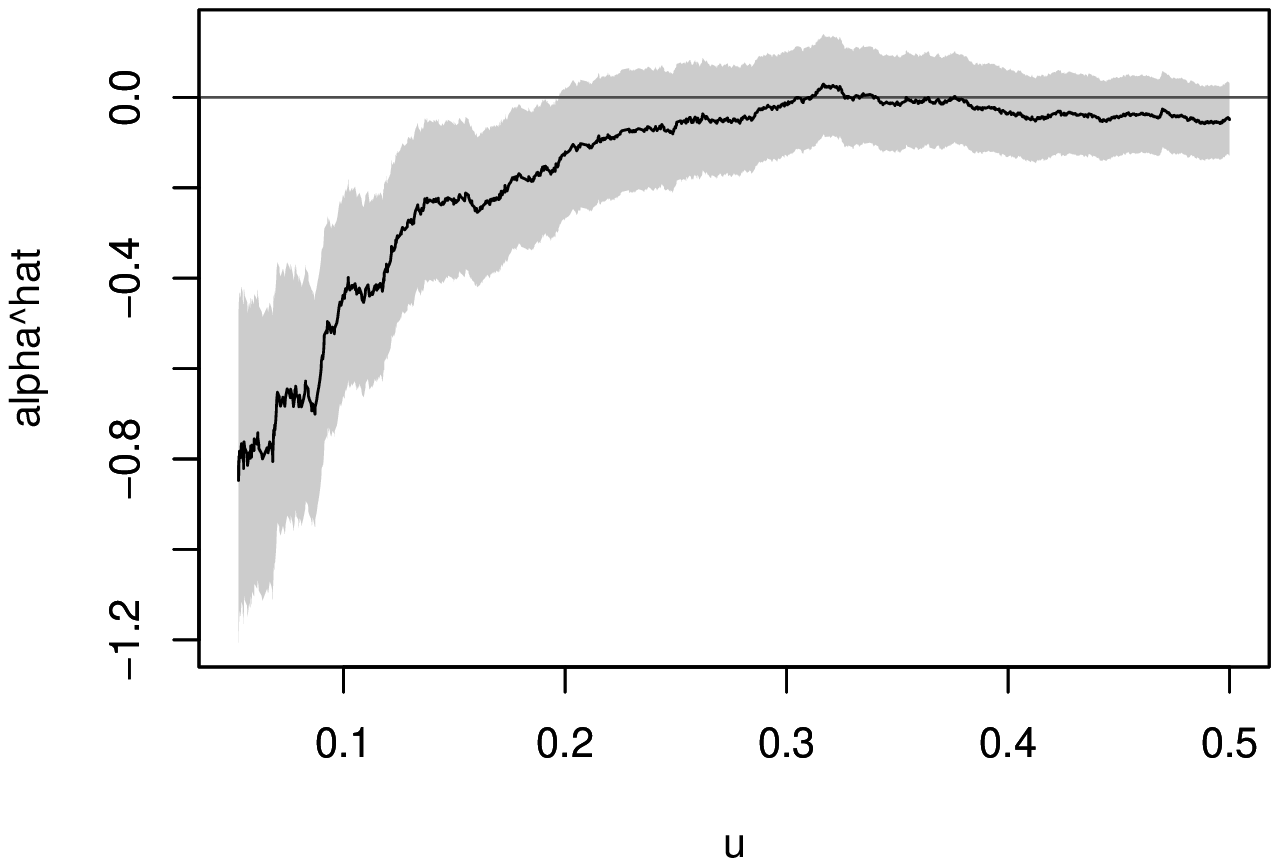} & 
\includegraphics[width=7cm,height=4.97cm]{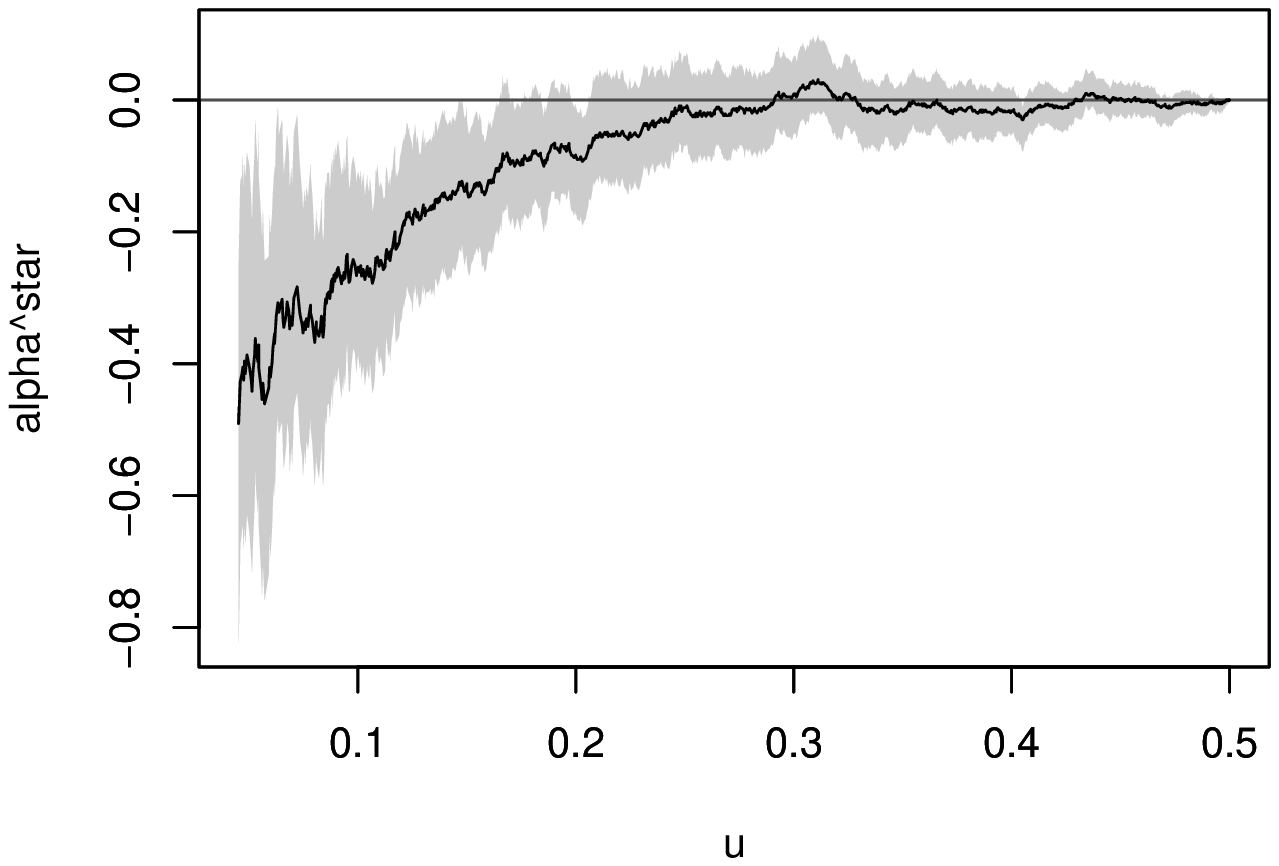} \\
\hspace{0.9cm} (d) & \hspace{0.9cm} (e)\vspace{0.7cm}
\end{tabular}
}
\end{tabular}
\caption{Plots of the sample $\{(u_{1i},u_{2i})\}_{i=1}^{2605}$ which the residuals are transformed into via the cumulative distribution functions of: (a) Student $t$-distribution and (b) empirical distribution.
(c) Plot of $- \varrho_K(a,u)$, a modified version of the measure (\ref{eq:krupskii}) of \citet{kru17}, with: $a(x)=x$ \skh{(solid, black), $a(x)=x^2$ (dashed, blue), and $a(x)=x^4$ (dotted, red).}
Plots of the proposed measure \skh{(black) and its \skh{asymptotic or bootstrap} 90\% confidence intervals (gray)} for: (d) $\hat{\alpha}(u)$ and (e) $\hat{\alpha}^* (u)$.
}\label{fig:example}
\end{center}
\end{figure}
Figure \ref{fig:example}(a) plots the sample $ \{(u_{1i},u_{2i}) \}$ which the residuals are transformed into via the cumulative distribution functions of Student $t$.
The values of $\hat{\alpha}(u)$ calculated from the sample and their 90\% asymptotic confidence intervals (\ref{eq:interval}) are displayed in Figure \ref{fig:example}(d).
In Figure \ref{fig:example}(d), the minimum value of $u$ is defined as \skd{$u_{\rm min}= \min \{ u \in (0,0.5] ;  T_L(u) ,T_U(u) \geq 30 \} \simeq 0.0526$ in order that the asymptotic theory is applicable.}

For the calculation of $\hat{\alpha}^*(u)$, we use the empirical distribution functions (\ref{eq:f_emp}) to transform the residuals $\{(x_{1i},x_{2i})\}$ into the copula sample $\{ (u_{1i},u_{2i}) \}$.
The transformed sample is displayed in Figure \ref{fig:example}(b).
The values of $\hat{\alpha}^*(u)$ calculated from the sample are plotted in Figure \ref{fig:example}(d).
\skh{The same frame also plots the 90\% confidence intervals based on $999$ resamples of size $2605$ using the basic bootstrap method.}
The minimum value of $u$ in the plot is $u^*_{\rm min}=  \min \{ u \in (0,0.5] ;  T_L^*(u) ,T_U^*(u) \geq 30 \} \simeq 0.0453$.


Figure \ref{fig:example}(a) and (b) suggest that there are more observations in the lower-left $[0,u]^2$ tail than the upper-right $[1-u,1]^2$ one for $u \simeq 0.1$.
However it does not seem immediately clear from these data plots whether there is significant difference between the two tail probabilities for $u \simeq 0.1$ as well as for general $u$.
To solve this problem, Figure \ref{fig:example}(d) and (e) showing the values of $\alpha(u)$ and $\alpha^*(u)$, respectively, are helpful.
Indeed Figure \ref{fig:example}(d) and (e) suggest that $\alpha(u)$ and $\alpha^*(u)$ are negative in most areas of the domain of $u$, \skh{suggesting} that the lower tail probability is greater than the upper one for most values of $u \in (u_{\rm min},0.5]$.
In particular the two frames imply the general tendency that, for $u \leq 0.2$, $\alpha(u)$ and $\alpha^*(u)$ decrease \skh{with $u$}.
The \skh{asymptotic and bootstrap} 90\% confidence intervals do not include $0$ for $u \leq 0.19$ in Figure \ref{fig:example}(d) and for $u \leq 0.14$ in Figure \ref{fig:example}(e).
Hence, when considering the tests $H_0:\alpha(\skc{u_0})=0 $ against $H_1:\alpha(\skc{u_0}) \neq 0$ for a fixed value of $\skc{u_0} \in [u_{\rm min},0.14]$ based on the two 90\% confidence intervals, both tests reject the null hypothesis at a significance level of \skh{0.1}.
This implies that the lower $[0,\skc{u_0})^2$ tail probability is significantly greater than the upper $(1-\skc{u_0},1]^2$ one for $\skc{u_0} \in [u_{\rm min},0.14]$.
On the other hand, when $\skc{u_0}$ is greater than 0.21, both 90\% confidence intervals include 0 and therefore each of the \skh{tests for a nominal size of 0.1} accepts the null hypothesis $H_0$.
There is disagreement in conclusions between the tests based on the two \skh{90\%} confidence intervals in some areas of $u$ in $(0.14,0.21]$.

As seen in the discussion above, both $\hat{\alpha}(u)$ and $\hat{\alpha}^*(u)$ show similar tendencies in general.
Actually, the two data plots given in Figure \ref{fig:example}(a) and \skh{(b) look} similar at the first glance.
However Figure \ref{fig:example}(d) and (e) reveal that there are some differences between $\hat{\alpha}(u)$ and $\hat{\alpha}^*(u)$.
For example, the values of $\hat{\alpha}(u)$ are generally smaller than those of $\hat{\alpha}^*(u)$ for $u \in (0,0.1]$.
Also the bootstrap confidence intervals of $\hat{\alpha}^*(u)$ are narrower than the asymptotic confidence intervals of $\hat{\alpha}(u)$ for large $u$.

Apart from the tests based on pointwise confidence intervals given in Figure \ref{fig:example}(d) and (e), we carry out a different \skh{test for a nominal size of 0.1} based on the test statistic in Theorem \ref{thm:test}.
We test $H_0: \alpha (u) = 0 $ against $H_1: \alpha (u) \neq 0 $ for $\{ u; u= u_{\rm min} + (0.1-u_{\rm min}) j / 10, \ j=0,\ldots,10\}$.
The test statistic is $T=\bm{a}^T \hat{\bm \Sigma}^{-1} \bm{a} \simeq 26.64$ with $\mathbb{P} (T > 26.64) \simeq 0.0052 \ll 0.1 $.
Therefore we reject the null hypothesis that the lower $[0,u]^2$ tail and upper $[1-u,1]^2$ tail are symmetric for the 11 equally spaced points of $u$ in $[u_{\rm min},0.1]$.

We apply other measures of tail asymmetry to the copula sample displayed in Figure \ref{fig:example}(a).
The measure (\ref{eq:rosco_joe}) of \citet{ros} is calculated as $\hat{\varsigma}_3 \simeq 0.0326$.
Another measure we consider here is a modified version \skc{of} \citeauthor{kru17}'s \citeyearpar{kru17} measure (\ref{eq:krupskii}), namely, $- \varrho_K(a,u)$.
This modification is made to interpret the sign of the measure in the same manner as in that of ours.
Figure \ref{fig:example}(c) displays the estimates of $-\varrho_K(a,u)$ with respect to $u$ for \skh{the} three specific functions of $a$.
The three curves of the modified measure  $-\varrho_K(a,u)$ agree that \skf{there is stronger} correlation in the lower $[0,u]^2$ tail than the upper $[1-u,1]^2$ one for $u \geq 0.12$.
This is somewhat similar to the result based on our measure as well.
For $a(x)=x^4$, the correlation coefficient in the lower $[0,u]^2$ tail is greater than that in the upper $[1-u,1]^2$ tail for any $u \in [0,0.5]$.

Finally, we summarize the results of the analysis of stock daily return data.
The results based on the proposed measures suggest that the lower $[0,u]^2$ tail probability is greater than the upper $[1-u,1]^2$ one for most values of $u \in (u_{\rm min},0.5]$, where $u_{\rm min}  \simeq 0.0526$.
In particular, there is significant difference between the lower and upper tail probabilities for $u < 0.14$.
\ske{From the economic perspective, this result implies that the joint plunging probability is higher than the joint soaring probability with the threshold $u<0.14$.}
Therefore it is recommended to use a copula with asymmetric tails for \skh{an appropriate modeling of the residuals} of the daily return data appropriately.
\ske{The three cases of the measure of \citet{kru17} agree} that, for $u  > 0.12$, there is stronger correlation in the lower $[0,u]^2$ tail than in the upper $[1-u,1]^2$ one.
}

\sk{
\section{Discussion} \label{sec:discussion}
\skh{In this paper we have proposed a copula-based measure of asymmetry between the lower and upper tail probabilities.
It has been seen that the proposed measure has some properties which are desirable as a measure of tail asymmetry.
Sample analogues of the proposed measure have been presented, and statistical inference based on them, including point estimation, interval estimation and hypothesis testing, has been shown to be very simple.
The practical importance of the proposed measure has been demonstrated through statistical analysis of stock return data.}

This paper discusses \skh{a} measure for bivariate data.
However it is straightforward to extend the proposed bivariate measure to a multivariate one in a similar manner as in \citet{emb} \skh{and \citet{hof}.}
Let $(X_1, \ldots, X_d)$ be an $\mathbb{R}^d$-valued random vector with continuous univariate margins.
Then an extended measure of tail asymmetry for $(X_1, \ldots, X_d)$ is defined by
$$
{\bm A}(u) = \left(
    \begin{array}{cccc}
      \alpha_{11} (u) & \alpha_{12}(u) & \cdots & \alpha_{1d} (u) \\
      \alpha_{21} (u) & \alpha_{22}(u) & \cdots & \alpha_{2d} (u) \\
      \vdots & \vdots &  \ddots & \vdots \\
      \alpha_{d1}(u) & \alpha_{d2}(u) & \ldots & \alpha_{dd}(u)
    \end{array}
  \right),
$$
where $\alpha_{ij}$ is the proposed measure (\ref{eq:alpha}) of the random vector $(X_i,X_j) \ (i,j=1,\ldots,d)$.
The properties of each element of the measure ${\bm A}(u)$ are straightforward from the results of this paper.
It would be a possible topic for future work to investigate properties of this extended measure as a matrix \skh{and evaluate the values of the measure for multivariate copulas such as some examples of the vine copulas \citep{aas,cza}.}
}

\section*{Supplementary material}
Supplementary material contains the proofs of Lemma \ref{lem:t} and Theorems \ref{thm:tail_dep}--\ref{thm:test} and plots of random variates from the copulas discussed in Sections \ref{sec:asymmetric_copula} and \ref{sec:alternative}.

\section*{Acknowledgements}
The authors are grateful to Hideatsu Tsukahara for his valuable comments on the work.
Kato's research was supported by JSPS KAKENHI Grant Numbers JP17K05379 and JP20K03759.

%




\newpage

\def\theequation{S\arabic{equation}}
\setcounter{equation}{0}

\def\thelemma{S\arabic{lemma}}
\setcounter{lem}{0}

\def\thetheorem{S\arabic{theorem}}
\setcounter{thm}{0}

\def\thesection{S\arabic{section}}
\setcounter{section}{0}

\def\thefigure{S\arabic{figure}}
\setcounter{figure}{0}

\def\thedefinition{S\arabic{definition}}
\setcounter{defi}{0}

\def\thetable{S\arabic{table}}
\setcounter{table}{0}

\title{\textbf{\Large Supplementary Material for ``Copula-based measures of asymmetry between the lower and upper tail probabilities"} \vspace{0.2cm} \\
}
\author{{\scshape Shogo Kato\(^{*,a}\),\ Toshinao Yoshiba\(^{b,c}\)  and Shinto Eguchi\,\(^{a}\)} \vspace{0.5cm}\\
\textit{\(^{a}\) Institute of Statistical Mathematics} \\  
\textit{\(^{b}\) Tokyo Metropolitan University} \\
\textit{\(^{c}\) Bank of Japan}
}
\date{August 4, 2020}
\maketitle

The Supplementary Material is organized as follows.
Section \ref{sec:proofs} presents the proofs of Lemma \ref{lem:t} and Theorems \ref{thm:tail_dep}--\ref{thm:test} of the \skh{article}.
Section \ref{sec:rv} displays plots of random variates from Clayton copula, Ali-Mikhail-Haq copula and BB7 copula discussed in Sections \ref{sec:asymmetric_copula} and \ref{sec:alternative} of the \skh{article}.

\section{Proofs} \label{sec:proofs}

\sk{
\subsection{Proof of Theorem \ref{thm:tail_dep}} \label{sec:tail_dep}

\begin{proof}
It follows from the expression (\ref{eq:lambda_u2}) that
$$
\lambda_U =  \lim_{u \uparrow 1} \frac{\overline{C}(u,u)}{1-u} = \lim_{u \downarrow 0} \frac{\overline{C}(\bar{u},\bar{u})}{u}.
$$
This result and \skd{Proposition \ref{thm:copula}} imply that
\begin{align*}
\alpha(0) & = \lim_{u \downarrow 0} \log \left( \frac{\overline{C}(\bar{u},\bar{u})}{C(u,u)} \right) = \log \left( \lim_{u \downarrow 0} \frac{\overline{C}(\bar{u},\bar{u})/u}{C(u,u)/u} \right) \skc{ = \log \left( \frac{\lambda_U}{\lambda_L} \right).}
\end{align*}
The last equality holds because $\lambda_U$ and $\lambda_L$ exist and either of $\lambda_U$ and $\lambda_L$ is not equal to zero.
\end{proof}
}

\sk{
\subsection{Proof of Theorem \ref{thm:tail_order}} \label{sec:tail_order}
\begin{proof}
If follows from the assumption that there exists a slowly varying function $\ell_L(u)$ such that $ C(u,u) \sim u^{\kappa_L} \ell_L(u) $ as $u \rightarrow 0$.
Similarly, there exists a slowly varying function $\ell_U(u)$ such that $ \overline{C}(\bar{u},\bar{u}) \sim u^{\kappa_U} \ell_U(u) $ as $u \rightarrow 0. $
Therefore
\begin{align*}
\alpha (0) & = \lim_{u \downarrow 0} \log \left( \frac{\overline{C}(\bar{u},\bar{u})}{C(u,u)} \right) = \lim_{u \downarrow 0} \log \left( u^{\kappa_U-\kappa_L} \frac{\ell_U(u)}{\ell_L(u)} \right) \\
 & = \left\{
\begin{array}{ll}
\infty , & \kappa_U > \kappa_L, \\
-\infty, & \kappa_U < \kappa_L.
\end{array}
\right.
\end{align*}
The last equality holds because $\ell_U(u)/\ell_L(u)$ is slowly varying.
If $\kappa_U=\kappa_L$ and either $\Upsilon_U \neq 0$ or $\Upsilon_L \neq 0$, then
$$
\alpha (0) = \lim_{u \downarrow 0} \log \left( \frac{\ell_U(u)}{\ell_L(u)} \right) = \log \left( \lim_{u \downarrow 0}  \frac{\ell_U(u)}{\ell_L(u)} \right)  = \log \left( \frac{\Upsilon_U}{\Upsilon_L} \right).
$$
\end{proof}
}

\sk{
\subsection{Proof of Theorem \ref{thm:alpha0_den}} \label{sec:alpha0_den}

\begin{proof}
\skd{Proposition \ref{thm:copula}} implies that $\alpha(0)$ can be expressed as
$$
\alpha(0) = \lim_{u \downarrow 0} \log \left( \frac{\overline{C}(\bar{u},\bar{u})}{C(u,u)} \right) = \log \left( \lim_{u \downarrow 0} \frac{\overline{C}(\bar{u},\bar{u})}{C(u,u)} \right).
$$
Since $\lim_{u \downarrow 0} d C(u,u) /du = \lim_{u \downarrow 0} d \overline{C}(\bar{u},\bar{u}) / du  = 0$, the l'H\^opital's rule is applicable to the last expression of the equation above.
Hence we have
\begin{align*}
\alpha(0) & = \log \left( \lim_{u \downarrow 0} \frac{d^2\overline{C}(\bar{u},\bar{u}) /du^2 }{d^2 C(u,u)/du^2} \right) =  \log \left( \lim_{u \downarrow 0} \frac{c(1-u)}{c(u)} \right)
\end{align*}
as required.
\end{proof}
}

\sk{
\subsection{Proof of Lemma \ref{lem:t}} \label{sec:t}

\begin{proof}
It is straightforward to see that $\mathbb{E} \left[ T_L(u) \right]$ and ${\rm var} \left[ T_L(u) \right]$ can be calculated as
\begin{align*}
\mathbb{E} \left[ T_L(u) \right] & = \mathbb{E} \left[ \frac{1}{n} \sum_{i=1}^n \bm{1} (U_{1i} \leq u, U_{2i} \leq u) \right] = \frac{1}{n} \sum_{i=1}^n \mathbb{E} \left[ \bm{1} (U_{1i} \leq u, U_{2i} \leq u) \right] \\
 & = \frac{1}{n} \cdot n C_u  = C_u,
\end{align*}
\begin{align*}
{\rm var} \left[ T_L(u) \right] & = {\rm var} \left[ \frac{1}{n} \sum_{i=1}^n \bm{1} (U_{1i} \leq u, U_{2i} \leq u) \right] \\
 & = \frac{n}{n^2} {\rm var} \left[ \bm{1} (U_{11} \leq u, U_{21} \leq u) \right] \\
  & = \frac{1}{n} \left( \mathbb{E} \left[ \bm{1}^2 (U_{11} \leq u, U_{21} \leq u) \right] -  \left\{ \mathbb{E} \left[ \bm{1} (U_{11} \leq u, U_{21} \leq u) \right]  \right\}^2 \right) \\
   & = \frac{1}{n} \left( C_u - C_u^2 \right) = \frac{1}{n} C_u (1-C_u).
\end{align*}
Noting that $\mathbb{E} \left[ \bm{1} (U_{11} > 1-u, U_{21} > 1-u) \right] = \overline{C}_{\bar{u}} $, the other expectation and variance, namely, $\mathbb{E} \left[ T_U(u) \right]$ and ${\rm var} \left[ T_U (u) \right]$, can be calculated in a similar manner.

Consider
$$
{\rm cov} \left[ T_L(u) , T_L(v) \right] = \mathbb{E} \left[ T_L(u) T_L(v) \right] -  \mathbb{E} \left[ T_L(u) \right] \mathbb{E} \left[ T_L(v) \right].
$$
The first term of the left-hand side of the equation above is
\begin{align*}
\mathbb{E} \left[ T_L(u) T_L(v) \right] = & \ \frac{1}{n^2} \mathbb{E} \left[ \sum_{i=1}^n \bm{1} (U_{1i} \leq u, U_{2i}  \leq u) \sum_{j=1}^n  \bm{1} (U_{1j} \leq v, U_{2j} \leq v)  \right]  \\
 = & \ \frac{1}{n^2} \sum_{i,j=1}^n \mathbb{E} \left[ \bm{1} (U_{1i} \leq u, U_{2i}  \leq u) \bm{1} (U_{1j} \leq v, U_{2j} \leq v)  \right] \\
  = & \ \frac{1}{n^2} \sum_{i=1}^n \mathbb{E} \left[ \bm{1} (U_{1i} \leq u , U_{2i}  \leq u ) \bm{1} (U_{1i} \leq v , U_{2i}  \leq v )\right]  \\
 & + \frac{1}{n^2} \sum_{i \neq j} \mathbb{E} \left[ \bm{1} (U_{1i} \leq u, U_{2i}  \leq u) \bm{1} (U_{1j} \leq v, U_{2j} \leq v) \right] \\
 = & \ \frac{1}{n^2} \sum_{i=1}^n \mathbb{E} \left[ \bm{1} (U_{1i} \leq u \wedge v, U_{2i}  \leq u \wedge v) \right]  \\
 & +  \frac{1}{n^2} \sum_{i \neq j} \mathbb{E} \left[ \bm{1} (U_{1i} \leq u, U_{2i}  \leq u) \right] \mathbb{E} \left[ \bm{1} (U_{1j} \leq v, U_{2j} \leq v) \right] \\
 = & \ \frac{1}{n^2} \left\{ n C_{u \wedge v} + n(n-1) C_u C_v \right\} = \frac{1}{n} C_{u \wedge v} \left\{ 1 + (n-1) C_{u \vee v} \right\}.
\end{align*}
Therefore we have
$$
{\rm cov} \left[ T_L(u) , T_L(v) \right] = \frac{1}{n} C_{u \wedge v} \left\{ 1 + (n-1) C_{u \vee v} \right\} - C_u C_v = \frac1n C_{u\wedge v} ( 1 - C_{u \vee v} ).
$$
Similarly, ${\rm cov} \left[ T_U(u) , T_U(v) \right]$ can be calculated.
The other covariance $ {\rm cov} [ T_L(u) , T_U(v) ] $ can also be obtained via a similar approach, but notice that
\begin{align*}
\mathbb{E} \left[ T_L(u) T_U(v) \right] = & \ \frac{1}{n^2} \sum_{i=1}^n \mathbb{E} \left[ \bm{1} (U_{1i} \leq u , U_{2i}  \leq u ) \bm{1} (U_{1i} > 1-v , U_{2i}  > 1-v )\right]  \\
 & + \frac{1}{n^2} \sum_{i \neq j} \mathbb{E} \left[ \bm{1} (U_{1i} \leq u, U_{2i}  \leq u) \bm{1} (U_{1j} > 1-v, U_{2j} > 1-v) \right] \\
 = & \ 0 + \frac{n(n-1)}{n^2} C_u \overline{C}_{\bar{v}} = \frac{n-1}{n} C_u \overline{C}_{\bar{v}}.
\end{align*}
The second equality holds because $0 < u,v \leq 0.5$.
Thus
$$
{\rm cov} \left[ T_L(u), T_U(v) \right] = \mathbb{E} \left[ T_L(u) T_U(v) \right] - \mathbb{E} \left[ T_L(u) \right] \mathbb{E} \left[ T_U(v) \right] =  -\frac{1}{n} C_u \overline{C}_{\bar{v}}.
$$
\end{proof}
}

\subsection{Proof of Theorem \ref{thm:gauss_con}} \label{sec:gauss_con}

\begin{proof}
Without loss of generality, assume $ 0 < u \leq  v \leq 0.5 $.
Let
$$
\hat{\bm{\beta}} =(\hat{\beta}_1,\hat{\beta}_2, \hat{\beta}_3, \hat{\beta}_4)^T = ( T_L(u),T_U(u),T_L(v),T_U(v) )^T,
$$
$$
\bm{\beta} = (\beta_1,\beta_2,\beta_3,\beta_4)^T = (C(u,u),\overline{C}(\bar{u},\bar{u}), C(v,v),\overline{C}(\bar{v},\bar{v}) )^T,
$$
$$
\bm{\Sigma_{\beta}} = (\sigma_{\beta i j} )_{i,j}, \quad \sigma_{\beta i j} = n \cdot {\rm cov} (\hat{\beta}_i , \hat{\beta}_j).
$$
Then it follows from Lemma \ref{lem:t} and the central limit theorem that
$$
\sqrt{n} ( \hat{\bm{\beta}} - \bm{\beta})  \xrightarrow{d} N ( \bm{0} , \bm{ \Sigma_{\beta}}) \quad \mbox{as } n \rightarrow \infty .
$$
Define
$$
h (\bm{\beta}) =  \left(
\begin{array}{c}
\log (\beta_2 / \beta_1 ) \\
\log (\beta_4 / \beta_3 )
\end{array}
\right) =
\left(
\begin{array}{c}
\alpha (u)  \\
\alpha (v)
\end{array}
\right).
$$
Applying the delta method, we have
$$
\sqrt{n} \left\{ h(\bm{\beta}) - h(\hat{\bm{\beta}}) \right\}  \xrightarrow{d} N \left( 0 , \nabla h(\bm{\beta})^T \bm{ \Sigma_{\beta}} \nabla h(\bm{\beta}) \right) \quad \mbox{as } n \rightarrow \infty ,
$$
where
$$
\nabla h (\bm{\beta}) = \left(
\begin{array}{cc}
\frac{\partial }{\partial \beta_1} \log ( \beta_2 / \beta_1 ) & \frac{\partial}{\partial \beta_1} \log (\beta_4/\beta_3) \\
\vdots & \vdots \\
\frac{\partial }{\partial \beta_4} \log (\beta_2/\beta_1) & \frac{\partial}{\partial \beta_4} \log (\beta_4/\beta_3) \\
\end{array}
\right) = \left(
\begin{array}{cc}
-1/\beta_1 & 0 \\
1/\beta_2 & 0 \\
0 & -1/\beta_3 \\
0 & 1/\beta_4
\end{array}
\right).
$$
\skc{The} asymptotic variance can be calculated as
\begin{eqnarray}
\nabla h(\bm{\beta})^T \bm{\Sigma_{\beta}} \nabla h(\bm{\beta}) &=& \left(
\begin{array}{cc}
\frac{\sigma_{11}}{\beta_1^2} - \frac{2 \sigma_{12} }{ \beta_1 \beta_2 } + \frac{ \sigma_{22} }{ \beta_2^2}   & \frac{\sigma_{13}}{\beta_1 \beta_3} - \frac{\sigma_{23}}{ \beta_2 \beta_3 } - \frac{\sigma_{14} }{ \beta_1 \beta_4 } + \frac{ \sigma_{24} }{ \beta_2 \beta_4 } \\
\frac{\sigma_{13}}{\beta_1 \beta_3} - \frac{\sigma_{23}}{ \beta_2 \beta_3 } - \frac{\sigma_{14} }{ \beta_1 \beta_4 } + \frac{ \sigma_{24} }{ \beta_2 \beta_4 } &  \frac{\sigma_{33}}{\beta_3^2} - \frac{2 \sigma_{34} }{ \beta_3 \beta_4 } + \frac{ \sigma_{44} }{ \beta_4^2}
\end{array}
\right)  \nonumber \\
 & =& \left(
\begin{array}{cc}
\frac{C(u,u)+ \overline{C}(\bar{u},\bar{u}) }{ C(u,u) \cdot \overline{C} (\bar{u},\bar{u}) } & \frac{C(v,v)+ \overline{C}(\bar{v},\bar{v}) }{ C(v,v) \cdot \overline{C} (\bar{v},\bar{v}) } \\
\frac{C(v,v)+ \overline{C}(\bar{v},\bar{v}) }{ C(v,v) \cdot \overline{C} (\bar{v},\bar{v}) } & \frac{C(v,v)+ \overline{C}(\bar{v},\bar{v}) }{ C(v,v) \cdot \overline{C} (\bar{v},\bar{v}) }
\end{array}
 \right) \nonumber \\
 & =& \skc{ \left(
\begin{array}{cc}
\sigma(u,u) & \sigma (u,v) \\
\sigma(u,v) & \sigma (v,v)
\end{array}
\right) . \label{eq:asy_var} }
\end{eqnarray}
The case $0 < v < u \leq 0.5 $ can be calculated in the same manner.
Then, for any $0 < u,v \leq 0.5$, it follows that, as $n \rightarrow \infty$, $ (\mathbb{A}_n(u),\mathbb{A}_n(v)) (=\sqrt{n} \{ h(\bm{\beta})-h(\hat{\bm{\beta}})\})$ converges weakly to the two-dimensional Gaussian distribution with mean 0 and the covariance matrix \skc{(\ref{eq:asy_var})}.
\skd{Weak convergence of $ (\mathbb{A}_n(u_1), \ldots \mathbb{A}_n(u_m))$ to an $m$-dimensional centered Gaussian distribution for $u_1,\ldots,u_m \in (0,0.5] \ (u_i \neq u_j, \ i \neq j)$ can be shown in a similar manner.}
Therefore $\{ \mathbb{A}_n(u) \, | \, 0 < u \leq 0.5 \}$ converges weakly to a centered Gaussian process with covariance function $\sigma(u,v)$ as $n \rightarrow \infty$.
\end{proof}

\sk{
\subsection{Proof of Theorem \ref{thm:test}} \label{sec:test}

\begin{proof}
Theorem \ref{thm:gauss_con} implies that $\bm{a}$ converges weakly to an $m$-dimensional normal distribution $N(\bm{0} ,\bm{ \Sigma})$ as $n$ tends to infinity, where
$$
\bm{\Sigma} = \left(
    \begin{array}{cccc}
      \sigma^2(u_1) & \sigma (u_1,u_2) & \ldots & \sigma (u_1,u_m) \\
      \sigma (u_1,u_2) & \sigma^2 (u_2) & \ldots & \sigma (u_2,u_m) \\
      \vdots & \vdots & \ddots & \vdots \\
      \sigma (u_1,u_m) & \sigma (u_2,u_m) & \ldots & \sigma^2 (u_m)
    \end{array}
  \right),
$$
$\sigma^2(u_i) = \sigma(u_i,u_i)$, and $\sigma(u_i,u_j)$ is defined as in Theorem \ref{thm:gauss_con}.
Then we have $\bm{a}^T \bm{\Sigma}^{-1} \bm{a} \xrightarrow{d} \chi^2(m)$ as $n \rightarrow \infty$.
Since $T_L(u_j)$ and $T_U(u_j)$ are consistent estimators of $C(u_j,u_j)$ and $\overline{C}(\bar{u}_j,\bar{u}_j)$, respectively, it holds that, for any $(i,j)$, $\hat{\sigma}(u_i,u_j)$ converges in probability to $\sigma(u_i,u_j)$ as $n \rightarrow \infty$.
It then follows from Slutsky's theorem that $\bm{a}^T \hat{\bm \Sigma}^{-1} \bm{a} \xrightarrow{d} \chi^2(m)$ as $n \rightarrow \infty$.
\end{proof}
}

\skb{
\section{Plots of random variates from some existing copulas} \label{sec:rv}

Figure \ref{fig:rv} plots random variates from Clayton copula (\ref{eq:clayton}), Ali-Mikhail-Haq copula (\ref{eq:amh}) and BB7 copula (\ref{eq:bb7}) with some selected values of the parameter(s).
This figure is given to help an intuitive understanding of the distributions of those copulas discussed in the paper.

\begin{figure}
\begin{center}
\begin{tabular}{ccc}
\includegraphics[width=4.5cm,height=4.5cm]{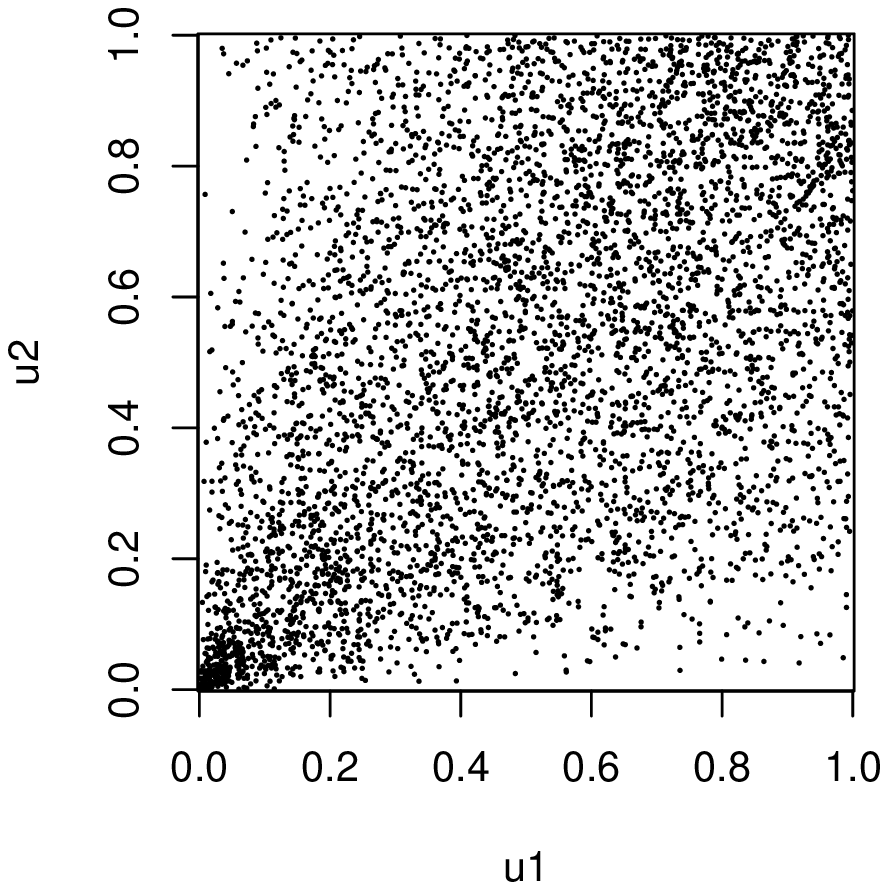} &
\includegraphics[width=4.5cm,height=4.5cm]{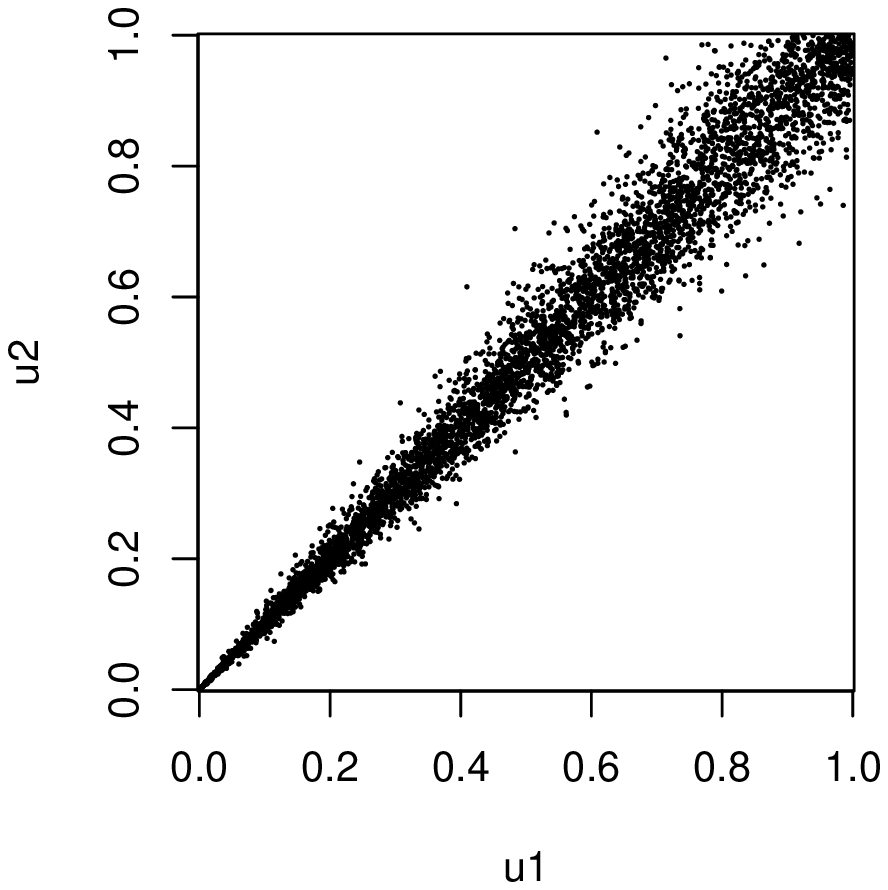} &
\includegraphics[width=4.5cm,height=4.5cm]{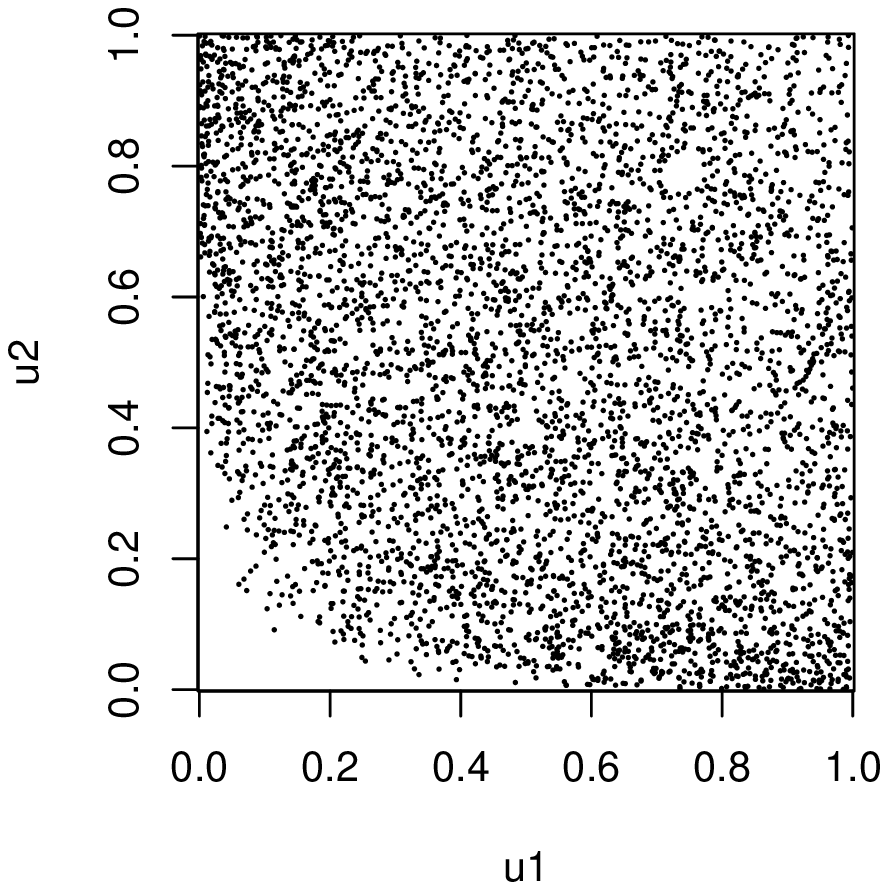} \\ 
\hspace{0.7cm} (a) & \hspace{0.7cm} (b) & \hspace{0.7cm} (c)\vspace{1cm}\\
\includegraphics[width=4.5cm,height=4.5cm]{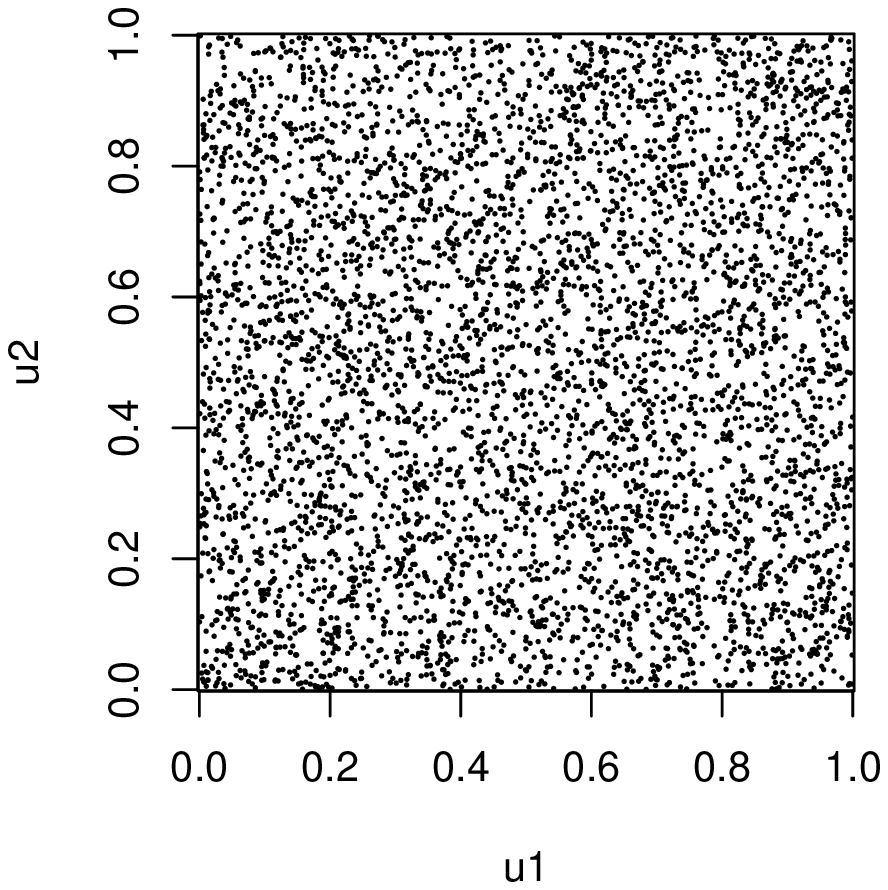} &
\includegraphics[width=4.5cm,height=4.5cm]{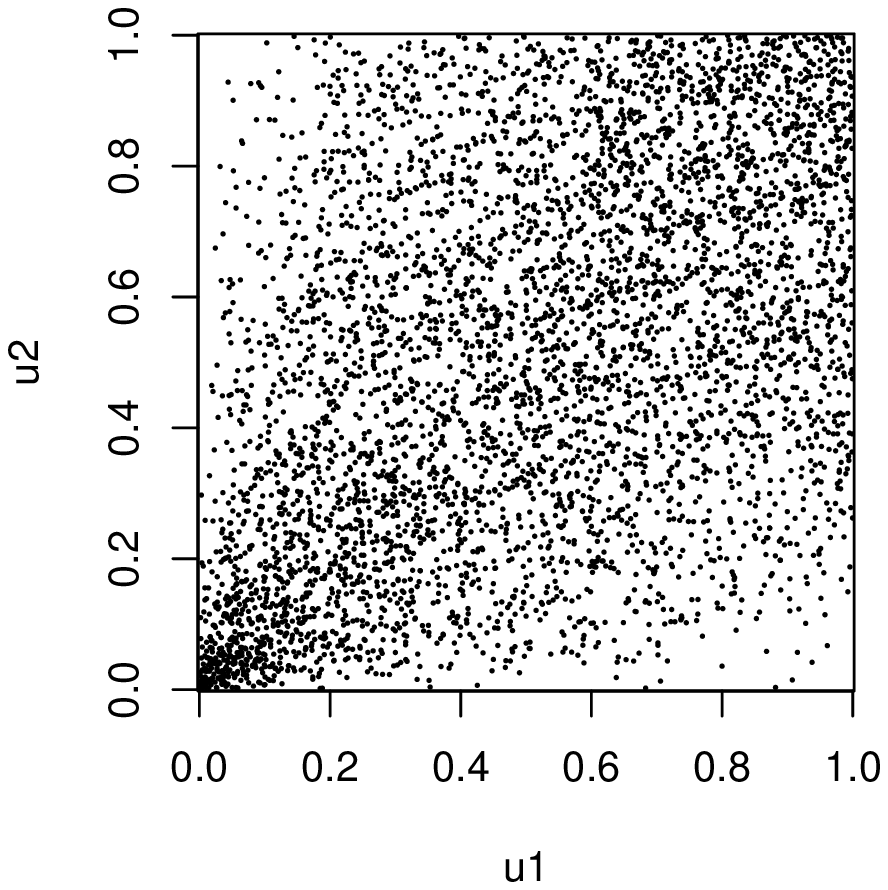} &
  \\ 
\hspace{0.7cm} (d) & \hspace{0.7cm} (e) & \vspace{1cm}\\
\includegraphics[width=4.5cm,height=4.5cm]{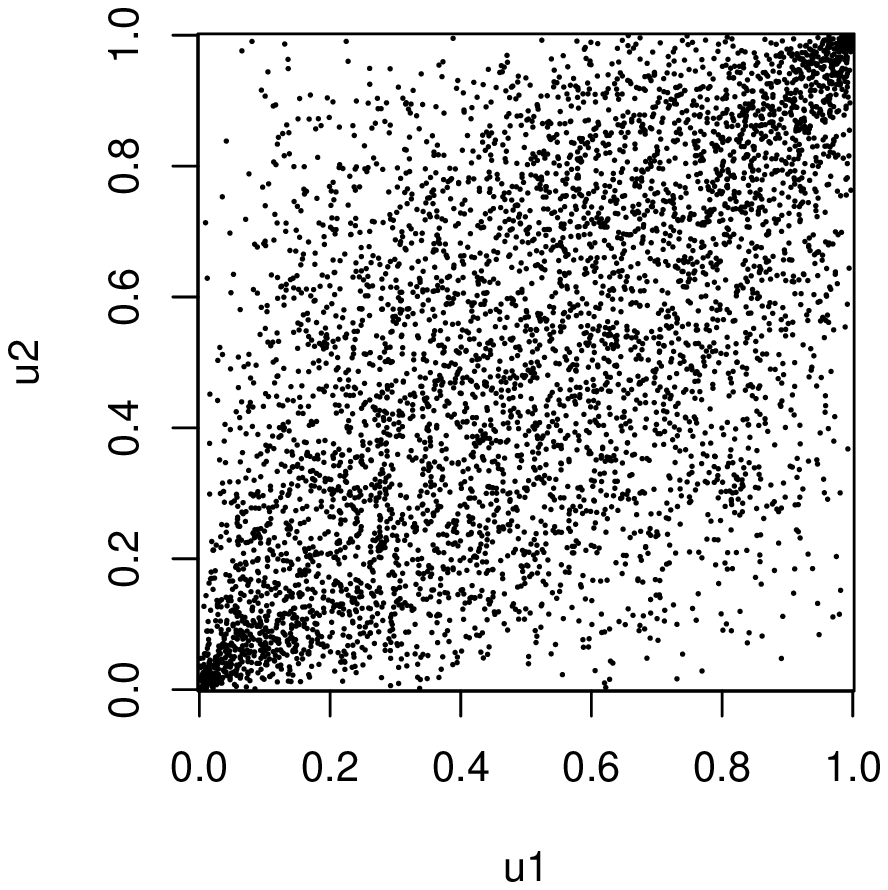} &
\includegraphics[width=4.5cm,height=4.5cm]{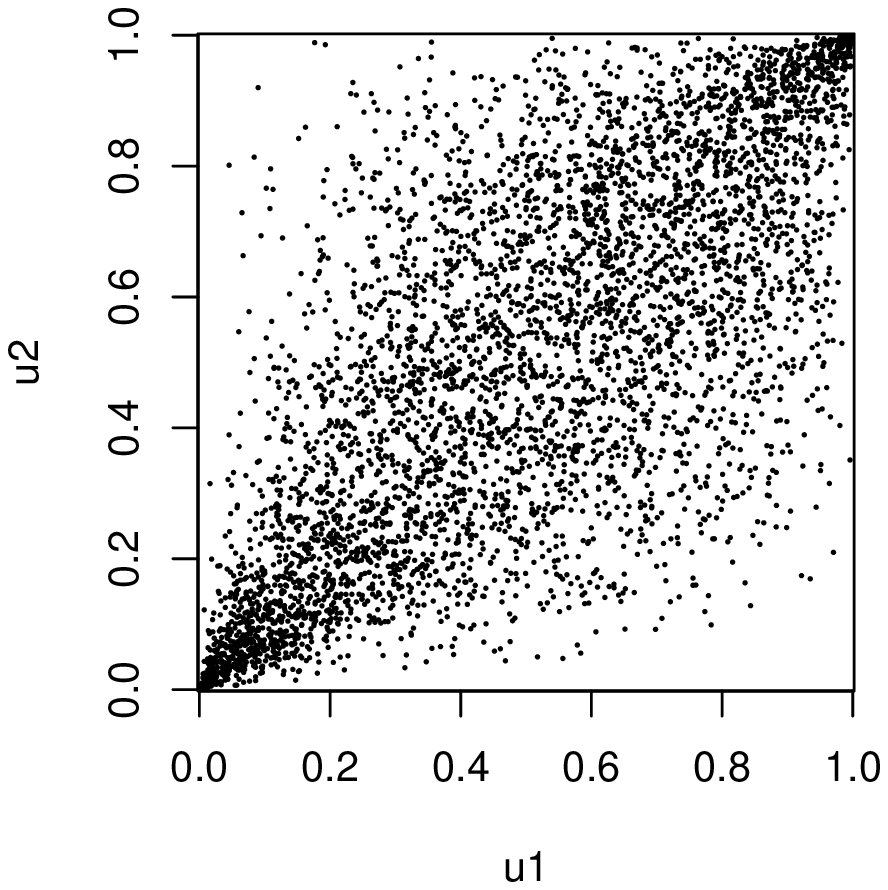} &
\includegraphics[width=4.5cm,height=4.5cm]{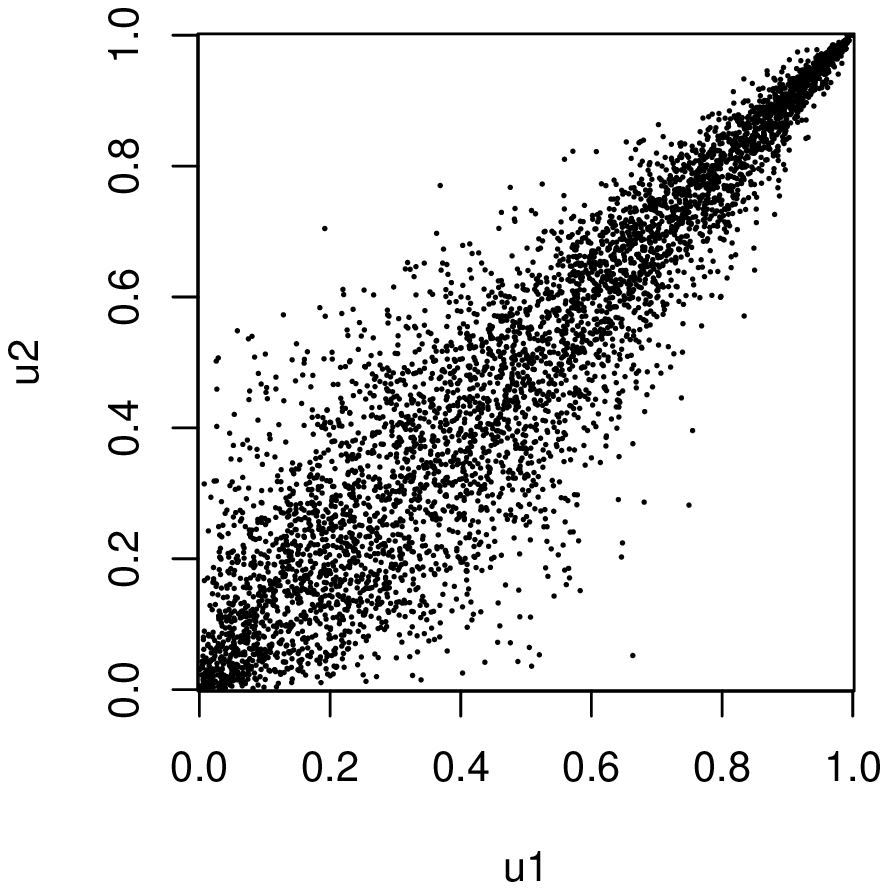} \\ 
\hspace{0.7cm} (f) & \hspace{0.7cm} (g) & \hspace{0.7cm} (h)\vspace{0.7cm}
\end{tabular}
\caption{\sk{Plots of 5000 random variates from: Clayton copula (\ref{eq:clayton}) with (a) $\theta=1$, (b) $\theta=20$ and (c) $\theta=-0.3$; Ali-Mikhail-Haq copula (\ref{eq:amh}) with (d) $\theta=0.1$ and $(e)$ $\theta=1$; and BB7 copula (\ref{eq:bb7}) with (f) $(\delta,\theta)=(1,1.71)$, (g) $(\delta,\theta)=(1.94,1.71)$ and (h) $(\delta,\theta)=(1,7.27)$.
} }\label{fig:rv}
\end{center}
\end{figure}
}

\bibliographystyle{Chicago}

\bibliography{Bibliography-MM-MC}

\end{document}